\documentclass[manuscript,noblind]{geophysics}

\usepackage{epstopdf}
\usepackage{graphicx}
\usepackage{amsmath}
\usepackage{mathrsfs}
\usepackage{algorithm}
\usepackage{algorithmic}
\usepackage{setspace}
\usepackage{bm}
\usepackage{amsfonts}

\usepackage{tikz}
\usetikzlibrary{positioning, shapes.geometric}
\usepackage{caption}
\usepackage{array}
\usepackage{booktabs}
\usepackage{amsfonts}
\usepackage{amssymb}
\usepackage{comment}
\usepackage{ifthen}
\usepackage{color}
\usepackage{upgreek}

\begin{document}
	
	\title{Efficient scattering approach to seismic full-waveform inversion in anisotropic elastic media with variable density
	}
	
	\renewcommand{\figdir}{Fig}
	\renewcommand{\thefootnote}{\fnsymbol{footnote}} 
	\address{
		$^\mathbf{1}$Department of Earth Science, University of Bergen, \\
		Postboks 7803, 5020 Bergen, Norway \\
		$^\mathbf{2}$Center for Modeling of Coupled Subsurface Dynamics, University of Bergen, \\
		Postboks 7803, 5020 Bergen, Norway \\
		$^\mathbf{3}$NORCE Norwegian Research Centre AS, \\
		Postboks 22 Nygårdstangen, 5838 Bergen, Norway \\
		\footnotemark[1] Corresponding author. \\
		\footnotemark[2] E-mail: Kui.Xiang@uib.no
	}
	\author{Kui Xiang$^\mathbf{1}$\footnotemark[1]\footnotemark[2], Morten Jakobsen$^\mathbf{1,2}$, Ujjwal Shekhar$^\mathbf{2}$, Kjersti Solberg Eikrem$^\mathbf{3}$ and Geir Nævdal$^\mathbf{3}$}
	
	\righthead{Multi-parameter inverse scattering}
	
	\maketitle
	
	\newpage
	
	\begin{abstract}
		
		This paper introduces a novel matrix-free approach for full waveform inversion in anisotropic elastic media, incorporating density variation through the utilization of the distorted Born iterative method. This study aims to overcome the computational and storage challenges associated with the conventional matrix-based distorted Born iterative inversion method while accurately capturing the subsurface's anisotropic properties and density variations. An elastic integral equation is utilized to account for the anisotropic nature of elastic wave propagation, enabling more precise modeling of subsurface complexities. This integral equation is efficiently solved by a fast Fourier transform accelerated Krylov subspace method. 	Leveraging the integral equation with the distorted Born approximation, a linear relationship between the scattered wavefield and the model parameter perturbation is formulated for an integrated inversion scheme. To address the inherent ill-posedness of each linear inversion step, we formulate the normal equation with a regularization term. This is achieved by minimizing an objective function using the generalized Tikhonov method. Therefore, we can find an adequate solution for the inverse scattering problem by solving the normal equation. Following the physical interpretation of Green's function, the Fr{\'e}chet and adjoint operators within the normal equation can be employed in a matrix-free manner, allowing for significant improvement of the computational efficiency and memory demand without compromising accuracy. The proposed matrix-free full waveform inversion  framework is thoroughly validated through extensive numerical experiments on synthetic datasets, showcasing its ability to reconstruct complex anisotropic structures and accurately recover stiffness parameters and density. This study contributes to advancing seismic imaging techniques and holds promise for applications in various domains, including hydrocarbon exploration, geological hazard assessment, and geotechnical engineering. Integrating stiffness parameters and density variation within the matrix-free full waveform inversion  framework paves the way for a more comprehensive understanding of subsurface properties.
		
	\end{abstract}
	
	\section{Introduction}
	
	Seismic full waveform inversion (FWI) is one of the most powerful tools for inferring the properties and structure of the subsurface by iteratively minimizing the residuals between calculated and observed data \citep{tarantola1984inversion,tarantola2005inverse,virieux2009overview}. It can provide a high-resolution reconstructed model by using the full data content, including diving waves, reflections, and multiples. Depending on the chosen forward-modeling approach, FWI can be carried out either in the time domain \citep{tarantola1984inversion,mora1988elastic,bunks1995multiscale} or in the frequency domain \citep{pratt1990inverse,zhou1995acoustic,liao1996multifrequency}. 
	From the mathematical physics perspective, FWI can be categorized as an inverse scattering problem. \citet{weglein2003inverse} provides a comprehensive overview of the essential logical steps and the foundational mathematical-physics principles behind the inverse scattering series used in seismic exploration. The inverse scattering problem has been extensively explored in physics, engineering, and applied sciences \citep{pike2001scattering}. There exist several highly developed methods for solving linear or nonlinear inverse scattering problems \citep{weglein1981obtaining,stolt1980inversion,van1997contrast,abubakar2004iterative,innanen2010direct,jakobsen2012nonlinear,haffinger2013towards,jakobsen2016renormalized,osnabrugge2016convergent,malovichko2018acoustic,song2020study,van2021forward}. Therefore, adapting these methods for application in seismic FWI is promising.
	
	One of these methods is the so-called distorted Born iterative (DBI) method. It is based on the integral equation method and was initially proposed for electromagnetic inverse scattering \citep{wang1989iterative,chew1990reconstruction}. 
	Although the implementation and performance of these methods are different, \citet{remis2000equivalence} and \citet{oristaglio2012wavefield} reported that the DBI method is equivalent to the Gauss–Newton method in optimization. The concept underlying the DBI method involves approximating the nonlinear inverse scattering problem with a sequence of linear problems. Then, each linear problem is solved to update the model incrementally. The DBI method has several notable characteristics: (1) it requires discretization solely for the anomalous region, enhancing efficiency for time-lapse inversion \citep{malovichko2017approximate,huang2019target}; 
	(2) it can provide the Fr{\'e}chet derivative (sensitivity kernel) directly via the use of Green's function, making it suitable for computing the Hessian information, which is crucial for reducing crosstalk effects in multi-parameter inversion \citep{jakobsen2020transition};  
	and (3) when Green's function pertains to homogeneous background media, its computation can be expedited using the fast Fourier transform (FFT) technique \citep{beylkin2009fast}.
	In the geophysical community, \citet{jakobsen2015full} introduced a T-matrix variant of the DBI method for solving the nonlinear problems in seismic FWI. \citet{huang2019target} verified that the T-matrix based DBI method is naturally target-oriented and successfully implemented it to the time-lapse seismic waveform inversion of a target region. \citet{eikrem2019iterated} proposed a Bayesian version of the DBI method for FWI, in which they also estimate the uncertainty surrounding the maximum a posteriori solution using the iterated extended Kalman filter. The main disadvantages of the DBI method are its large computational cost and memory requirements. However, these disadvantages can be circumvented by implementing the Fr{\'e}chet derivative and its adjoint in a fast and efficient manner \citep{xiang2023matrix}, in which the matrix-vector multiplications are achieved by operation of the Green's function with FFT. 
	
	The Earth's subsurface is defined by an intricate combination of parameters, including seismic velocities (P-wave and S-wave), density, attenuation, and anisotropy. Each of these parameters affects seismic wave propagation in unique ways. With the development of geophysical exploration technology, simple reservoirs with straightforward geological features become increasingly rare. Instead, we face intricate oil and gas reservoirs characterized by complex subsurface structures, including intricate tectonics, stratigraphy, and lithology. The imaging requirements of these complex reservoirs are such that we cannot ignore the impact of multiple parameters on the inversion results. By updating multiple parameters concurrently, multiparameter FWI can provide more accurate and high-resolution subsurface models \citep{operto2013guided}. In light of significant development in computer technology over recent decades, multiparameter FWI has been widely studied in numerous publications \citep{brossier2009seismic,kohn2012influence,prieux2013multiparameter,lin2014acoustic,wang2017elastic}. Some publications have shown the results of multiparameter FWI in the anisotropic situation \citep{burridge1998multiparameter,lee2010frequency,warner2013anisotropic,kamath2016elastic,oh2016elastic,pan2016estimation,alkhalifah2016recipe,he2017analysis,rusmanugroho2017anisotropic,yang2019analysis}. 
	
	In the seismic full-waveform inversion algorithm utilizing the DBI method, \citet{jakobsen2020transition} generalized the DBI T-matrix method so that it can be used in conjunction with the elastodynamic equation for the anisotropic elastic medium. \citet{huang2020bayesian} extended the Bayesian framework of \citet{eikrem2019iterated}, which incorporated the DBI method and the iterated extended Kalman filter, to anisotropic elastic media. Although these implementations of the DBI method in seismic multi-parameter inversion obtained good inversion results, the computational cost and memory requirements remain substantial due to the matrix formulation employed. \citet{jakobsen2023seismic} generalized the matrix-free DBI method from mono-parameter (velocity) to multi-parameter (velocity and density) FWI and extended the implementation of the matrix-free DBI method to seismic inversion and medical imaging. In this paper, we further extend the matrix-free DBI method to FWI in elastic anisotropic media. Compared to previous studies, the approach proposed in this paper has several innovative aspects: (1) we only use a single displacement-related integral equation instead of two coupled integral equations for the particle displacement and the strain fields; (2) this is matrix-free for FWI in elastic anisotropic media; (3) the reconstruction of density is included in both theory and numerical examples; 
	(4) we employ finite differences to compute the derivative of the Green's function rather than the previous analytical formulation for the derivative of the Green's function, which offers a reduced computational cost compared to the analytical formulation; and (5) we extend the application of the DBI method for FWI in elastic anisotropic media from 2D to 3D. 
	
	This paper is organized as follows. We first provide the integral equation of the forward problem for elastic anisotropic media, which considers the 21 elastic parameters and the mass density. We then describe a fast forward solver, in conjunction with the Krylov subspace method and the FFT, for solving the integral equation, which will be used for simulating the synthetic data. Next, we derive the mathematical formulation of the elastic DBI method and provide the matrix-free expressions of the Fr{\'e}chet derivative and its adjoint operators. The normal equation required for solving the inverse problem is also described. Finally, we use 2D and 3D numerical examples to test the performance of our method for elastic FWI in the transversely isotropic (VTI) media with a vertical symmetry axis. 
\newpage

\section*{Methodology}
\subsection{Forward modeling}

The propagation of seismic waves within a heterogeneous anisotropic elastic medium occupying a bounded domain $ D \subset \mathbf{R}^{d} $, $ d = 2 $, 3, is governed by the elastodynamic wave equation \citep{cerveny2005seismic,jakobsen2020transition}:
\begin{equation}\label{Ewave}
	\omega^{2} \rho(\mathbf{x}) \mathbf{u}(\mathbf{x}, \omega) +\nabla \cdot \left[\mathbf{C}(\mathbf{x})  : \nabla \mathbf{u} (\mathbf{x}, \omega)\right]=-\mathbf{S}(\mathbf{x}, \omega), 
\end{equation}
where $ \rho \ $ is the mass density, $ \mathbf{C} $ is the elastic stiffness tensor,  $ \omega $ is the angular frequency, $ \mathbf{u} \subset \mathbf{R}^{d} $, is the displacement, and $\mathbf{S} \subset \mathbf{R}^{d} $ is the source. The double dot product of the fourth rank tensor $ \mathbf{C} $ and the second rank tensor $ \nabla \mathbf{u} $ is defined by the summation over the two repeated indices \citep{auld1973acoustic}. 
For clarity in notation, we will omit the $\omega$-dependency within the displacement $\mathbf{u}$, the source $ \mathbf{S} $ and the following Green's function. The displacement field $ \mathbf{u} $ can be expressed as an integral over all space \citep{morse1954methods,gubernatis1977formal,jakobsen2020transition}, often referred to as the source representation, given by:
 	\begin{equation}\label{u_int}
 \mathbf{u} (\mathbf{x}) = \int d \mathbf{x}^{\prime} \mathbf{G}\left(\mathbf{x}, \mathbf{x}^{\prime} \right)  \mathbf{S}\left(\mathbf{x}^{\prime}\right),
 \end{equation}
 where $ \mathbf{G}\left(\mathbf{x}, \mathbf{x}^{\prime} \right) $ is the Green's function, defined by
 \begin{equation}\label{Green}
 \omega^{2} \rho(\mathbf{x}) \mathbf{G}(\mathbf{x},\mathbf{x}^{\prime}) +\nabla \cdot \left[\mathbf{C}(\mathbf{x})  : \nabla \mathbf{G}(\mathbf{x},\mathbf{x}^{\prime})\right]=-\delta(\mathbf{x}-\mathbf{x}^{\prime}),
 \end{equation}
where the delta function $ \delta(\mathbf{x}-\mathbf{x}^{\prime}) $ represents a unit point source.
 
	Consider a homogeneous anisotropic elastic model described by stiffness parameters $ \mathbf{C}^{(0)} $ and density $ \rho^{(0)} $ and introduce the perturbations of elastic parameters and density, $ \delta\mathbf{C}^{(0)} $ and $ \delta\rho^{(0)} $, as follows:
	\begin{equation}\label{decom}	
		\mathbf{C}(\mathbf{x}) = \mathbf{C}^{(0)} + \delta\mathbf{C}^{(0)}(\mathbf{x}), \qquad
		\rho(\mathbf{x}) = \rho^{(0)} + \delta\rho^{(0)}(\mathbf{x}).
	\end{equation}  
Inserting \eqref{decom} into \eqref{Ewave}, we obtain 
    \begin{equation}\label{Ewave0}
    	\omega^{2} \rho^{(0)} \mathbf{u}(\mathbf{x}) +\nabla \cdot \left[\mathbf{C}^{(0)}  : \nabla \mathbf{u} (\mathbf{x})\right]=-\mathbf{S}(\mathbf{x}) - \omega^{2} \delta\rho^{(0)}\left(\mathbf{x}\right)\mathbf{u}(\mathbf{x}) - \nabla \cdot\left[\delta\mathbf{C}^{(0)} \left(\mathbf{x}\right) : \nabla \mathbf{u}(\mathbf{x})\right].
    \end{equation}
    If we treat the right-hand side of \eqref{Ewave0} as a new source term, then $ \mathbf{u}(\mathbf{x}) $ can be interpreted as the displacement in the homogeneous background model due to the new source.     Using the volume integral \eqref{u_int}, we can represent equation \eqref{Ewave0} in the integral form:
	\begin{equation}\label{u}
	\begin{aligned} 
		\mathbf{u}\left(\mathbf{x}\right) = 
		& \mathbf{u}^{(0)}\left(\mathbf{x}\right)+ \omega^{2} \int d \mathbf{x}^{\prime} \mathbf{G}^{(0)}\left(\mathbf{x}- \mathbf{x}^{\prime}\right) \delta\rho^{(0)}\left(\mathbf{x}^{\prime}\right) \mathbf{u}\left(\mathbf{x}^{\prime}\right) \\
		& + \int d \mathbf{x}^{\prime} \mathbf{G}^{(0)} \left(\mathbf{x} - \mathbf{x}^{\prime} \right) \nabla_{\mathbf{x}^{\prime}} \cdot \left[ \delta\mathbf{C}^{(0)}\left(\mathbf{x}^{\prime}\right) : \nabla_{\mathbf{x}^{\prime}} \mathbf{u}\left(\mathbf{x}^{\prime}\right)\right],
	\end{aligned}  
\end{equation}
	where 
	\begin{equation}\label{u0}
		\mathbf{u}^{(0)} (\mathbf{x}) = \int d \mathbf{x}^{\prime} \mathbf{G}^{(0)} \left(\mathbf{x} - \mathbf{x}^{\prime} \right)  \mathbf{S}\left(\mathbf{x}^{\prime}\right)
	\end{equation}
    is the displacement in the homogeneous background medium caused by $ \mathbf{S}\left(\mathbf{x}^{\prime}\right) $ and $ \mathbf{G}^{(0)}\left(\mathbf{x} - \mathbf{x}^{\prime} \right) $ is the Green's function for the homogeneous medium. The Green's function for a homogeneous medium can be derived from equation \eqref{Green} by substituting the actual medium with the homogeneous medium. The analytical expressions of 2D and 3D elastodynamic Green's function for homogeneous media are given in Appendix~\ref{op_Green}. 
    Since Green's function for the homogeneous medium is translation invariant, we present it as a function of the vector difference  $ \mathbf{x}-\mathbf{x}^{\prime} $. More details about the Green's function can be found in \citet{arfken1999mathematical} and \citet{cerveny2005seismic}.  
    
Equation \eqref{u} involves the spatial derivative on perturbations $ \delta\mathbf{C}^{(0)} $, which is not convenient for subsequent inversion tasks. Therefore, we need to rewrite equation \eqref{u} in a more suitable form \citep{cerveny2005seismic}. Using
\begin{equation}\label{dgv}
\begin{aligned} 
\mathbf{G}^{(0)} \left(\mathbf{x} - \mathbf{x}^{\prime} \right) \nabla_{\mathbf{x}^{\prime}} \cdot \left[ \delta\mathbf{C}^{(0)}\left(\mathbf{x}^{\prime}\right) : \nabla_{\mathbf{x}^{\prime}} \mathbf{u}\left(\mathbf{x}^{\prime}\right)\right] = &
\nabla_{\mathbf{x}^{\prime}} \cdot \left[ \mathbf{G}^{(0)} \left(\mathbf{x} - \mathbf{x}^{\prime} \right) \delta\mathbf{C}^{(0)}\left(\mathbf{x}^{\prime}\right) : \nabla_{\mathbf{x}^{\prime}} \mathbf{u}\left(\mathbf{x}^{\prime}\right) \right] \\
& - \nabla_{\mathbf{x}^{\prime}} \mathbf{G}^{(0)} \left(\mathbf{x} - \mathbf{x}^{\prime} \right) \cdot \left[ \delta\mathbf{C}^{(0)}\left(\mathbf{x}^{\prime}\right) : \nabla_{\mathbf{x}^{\prime}} \mathbf{u}\left(\mathbf{x}^{\prime}\right)\right],
\end{aligned}  
\end{equation}
inserting this into equation \eqref{u}, we obtain
\begin{equation}\label{u_2}
\begin{aligned} 
\mathbf{u}\left(\mathbf{x}\right) = 
& \mathbf{u}^{(0)}\left(\mathbf{x}\right)+ \omega^{2} \int d \mathbf{x}^{\prime} \mathbf{G}^{(0)}\left(\mathbf{x}- \mathbf{x}^{\prime}\right) \delta\rho^{(0)}\left(\mathbf{x}^{\prime}\right) \mathbf{u}\left(\mathbf{x}^{\prime}\right) \\
& - \int d \mathbf{x}^{\prime} \nabla_{\mathbf{x}^{\prime}} \mathbf{G}^{(0)} \left(\mathbf{x} - \mathbf{x}^{\prime} \right) \cdot \left[ \delta\mathbf{C}^{(0)}\left(\mathbf{x}^{\prime}\right) : \nabla_{\mathbf{x}^{\prime}} \mathbf{u}\left(\mathbf{x}^{\prime}\right)\right].
\end{aligned}  
\end{equation}
Note that the volume integral over the first term on the right hand of equation \eqref{dgv} has been transformed into a surface integral which is identical to zero under the assumption that the displacement field $ \mathbf{u} $ approaches zero at infinity. 

Equation \eqref{u_2} is the desired integral equation for our computational approach. \citet{shekhar2023integral} proposed utilizing this single integral equation \eqref{u_2} to solve for the displacement field instead of relying on two coupled integral equations as presented in \citet{jakobsen2020transition}. In the work presented by \citet{shekhar2023integral}, the derivative of the Green's function $ \nabla \mathbf{G}^{(0)} $ is obtained analytically, while the derivative of the displacement $ \nabla \mathbf{u} $  is computed using the finite difference method. In this paper, the derivative of Green's function and the derivative of the displacement are computed using the finite difference method. Once the perturbations $ \delta\mathbf{C}^{(0)} $ and $ \delta\rho^{(0)} $, the incident wavefield $ \mathbf{u}^{(0)} $, and the Green's function $ \mathbf{G}^{(0)} $ are known, equation \eqref{u_2} enables us to calculate the displacement $ \mathbf{u} $. To solve $ \mathbf{u} $ from \eqref{u_2}, we first define an operator as 
	\begin{equation}\label{A}
\begin{aligned}
\mathcal{A} \left(\cdot\right) = & \mathcal{I}\left(\cdot\right) - \omega^{2} \int d \mathbf{x}^{\prime} \mathbf{G}^{(0)}\left(\mathbf{x}- \mathbf{x}^{\prime}\right) \delta\rho^{(0)}\left(\mathbf{x}^{\prime}\right) \left(\cdot\right)\\ & + \int d \mathbf{x}^{\prime} \nabla_{\mathbf{x}^{\prime}} \mathbf{G}^{(0)} \left(\mathbf{x} - \mathbf{x}^{\prime} \right)  \cdot \left[ \delta\mathbf{C}^{(0)}\left(\mathbf{x}^{\prime}\right) : \nabla_{\mathbf{x}^{\prime}} \left(\cdot\right)\right],
\end{aligned}
\end{equation}
where $ \mathcal{I} $ is the identity operator. Substituting \eqref{A} into \eqref{u_2}, we have 
	\begin{equation}\label{axb}
\mathcal{A}\left\{\mathbf{u}\left(\mathbf{x}\right)\right\} = \mathbf{u}^{(0)}\left(\mathbf{x}\right).
\end{equation}
Equation \eqref{axb} is a linear equation in the form of $ \mathbf{A} \mathbf{x} =\mathbf{b} $, which can be iteratively solved using Krylov subspace methods, such as the conjugate gradient method, BiCGSTAB (biconjugate gradient stabilized) method, GMRES (generalized minimum residual) method, among others. In operator \eqref{A}, the Green's function and its derivative, when multiplied by their subsequent vectors, can be interpreted as circular convolutions. These circular convolutions can be computed efficiently by using the Fast Fourier Transform. The computations are written symbolically as
\begin{equation}\label{FFT}
\mathbf{G}^{(0)} \left\{\mathbf{v}\right\} = \mathrm{FFT}^{-1} \{\mathrm{FFT}\{\mathbf{G}^{(0)} \}\odot\mathrm{FFT} \{\mathbf{v}\}\},
\end{equation}
where $ \mathbf{v} $ can be any vector and $ \odot $ denotes  element-wise multiplication.
 
	\subsection{The distorted Born iterative method}
Let us consider a known heterogeneous reference model described by elastic parameters $ \mathbf{C}^{(b)}$ and density $ \rho^{(b)} $. The differences in the elastic parameters and density between the true and the reference model are defined as
	\begin{equation}\label{del}
		\delta \rho\left(\mathbf{x}\right) = \rho\left(\mathbf{x}\right) - \rho^{(b)}\left(\mathbf{x}\right), \qquad 
		\delta \mathbf{C}\left(\mathbf{x}\right) = \mathbf{C}\left(\mathbf{x}\right) - \mathbf{C}^{(b)}\left(\mathbf{x}\right). 
	\end{equation}
The discrepancy between the true displacement field $ \mathbf{u} $ in the true model and the reference displacement field $ \mathbf{u}^{(b)} $ in the reference model is defined as 
	\begin{equation}\label{delu}
		\delta \mathbf{u}\left(\mathbf{x}\right) = \mathbf{u}\left(\mathbf{x}\right) - \mathbf{u}^{(b)}\left(\mathbf{x}\right).
	\end{equation}
By replacing the homogeneous background model in equation \eqref{u_2} with the heterogeneous reference model, we have	
	\begin{equation}\label{u_3}
		\begin{aligned} 
			\mathbf{u}\left(\mathbf{x}\right) = 
			& \mathbf{u}^{(b)}\left(\mathbf{x}\right)+ \omega^{2} \int d \mathbf{x}^{\prime} \mathbf{G}^{(b)}\left(\mathbf{x},  \mathbf{x}^{\prime}\right) \delta\rho\left(\mathbf{x}^{\prime}\right) \mathbf{u}\left(\mathbf{x}^{\prime}\right) \\
			& - \int d \mathbf{x}^{\prime} \nabla_{\mathbf{x}^{\prime}} \mathbf{G}^{(b)} \left(\mathbf{x}, \mathbf{x}^{\prime} \right) \cdot \left[ \delta\mathbf{C}\left(\mathbf{x}^{\prime}\right) : \nabla_{\mathbf{x}^{\prime}} \mathbf{u}\left(\mathbf{x}^{\prime}\right)\right], 
		\end{aligned}  
	\end{equation} 
	where $ \mathbf{G}^{(b)} \left(\mathbf{x}, \mathbf{x}^{\prime} \right) $ is the Green's function for a heterogeneous background medium that can be numerically solved from equation \eqref{Green}. The heterogeneous Green's function $ \mathbf{G}^{(b)} \left(\mathbf{x}, \mathbf{x}^{\prime} \right) $ can also be solved from the Dyson equations in \citet{jakobsen2015full}. However, in this paper, we will employ operations only involving the homogeneous Green's function as an alternative to the direct use of the heterogeneous Green's function, thereby circumventing the need for storage and updates of the Green's function associated with the model update. 
	When performing inversion, the data residual of the displacement field can only be observed at the receiver position $\mathbf{r}$. By combining this information with equation \eqref{u_3} and \eqref{delu}, we obtain
	\begin{equation}\label{du2}
		\delta \mathbf{u}\left(\mathbf{r}\right) = 
		\omega^{2} \int d \mathbf{x}^{\prime} \mathbf{G}^{(b)}\left(\mathbf{r},\mathbf{x}^{\prime}\right) \delta \rho\left(\mathbf{x}^{\prime}\right) \mathbf{u}\left(\mathbf{x}^{\prime}\right) 
		-\int d \mathbf{x}^{\prime} \nabla_{\mathbf{x}^{\prime}} \mathbf{G}^{(b)}\left(\mathbf{r}, \mathbf{x}^{\prime}\right)
		\cdot \left[ \delta\mathbf{C}\left(\mathbf{x}^{\prime}\right) : \nabla_{\mathbf{x}^{\prime}} \mathbf{u}\left(\mathbf{x}^{\prime}\right)\right].
	\end{equation}
    In equation \eqref{du2}, it is evident that to solve for $ \delta \rho $ and $ \delta\mathbf{C} $, we require knowledge of the displacement $ \mathbf{u}\left(\mathbf{x}^{\prime}\right) $ within the unknown true model, which is unavailable to us. However, we know the heterogeneous background model and its corresponding displacement field. Therefore, we assume that the known heterogeneous background model is close to the unknown true model, which results in the reference displacement being close to the true displacement. This assumption allows equation \eqref{du2} to be approximated as
	\begin{equation}\label{du}
		\delta \mathbf{u}\left(\mathbf{r}\right) \approx 
		\omega^{2} \int d \mathbf{x}^{\prime} \mathbf{G}^{(b)}\left(\mathbf{r},\mathbf{x}^{\prime}\right) \delta \rho\left(\mathbf{x}^{\prime}\right) \mathbf{u}^{(b)}\left(\mathbf{x}^{\prime}\right) 
		-\int d \mathbf{x}^{\prime} \nabla_{\mathbf{x}^{\prime}} \mathbf{G}^{(b)}\left(\mathbf{r}, \mathbf{x}^{\prime}\right)
		\cdot \left[ \delta\mathbf{C}\left(\mathbf{x}^{\prime}\right) : \nabla_{\mathbf{x}^{\prime}} \mathbf{u}^{(b)}\left(\mathbf{x}^{\prime}\right)\right].		
	\end{equation}
	In the distorted Born iterative method method, the displacement residual $ \delta \mathbf{u} $, the background Green's function $ \mathbf{G}^{(b)} $, and the background displacement $ \mathbf{u}^{(b)} $ 	are computed first based on a known background model. After that, we can solve \eqref{du} for $ \delta \mathbf{C} $ and $ \delta \rho $, and then iteratively update the known background model.
	
    \subsection{Fr{\'e}chet derivative and adjoint operators}
    
    To further progress, we must decompose the stiffness tensor perturbations into a spatially invariant tensor structure part and a spatially variable scalar function for each independent model parameter. Thus, we decompose $ \delta \mathbf{C}\left(\mathbf{x}\right) $ as
	\begin{equation}\label{dC}
		\delta \mathbf{C} \left(\mathbf{x}\right) = \sum_{p=1}^{21} \mathbf{B}^{(p)} \delta m^{(p)}\left(\mathbf{x}\right),  
	\end{equation}
	where $ \delta m^{(p)}, \quad p = 1,2,...,21 $ are the model perturbations of the 21 independent elastic parameters,	and $ \mathbf{B} $ is the tensor field structure related to the position of the elastic parameter in the stiffness tensor \citep{jakobsen2020transition}. To ensure consistency with the representation of $ \delta m^{(p)} $ in terms of elastic parameters, we introduce the notation $ \delta m^{(0)} = \delta \rho $ for the density.  
	Inserting \eqref{dC} into \eqref{du}, we obtain
 \begin{equation}\label{duB}
 	\begin{aligned}
 		\delta \mathbf{u}\left(\mathbf{r}\right) \approx & 
 		\omega^{2} \int d \mathbf{x}^{\prime} \mathbf{G}^{(b)}\left(\mathbf{r},\mathbf{x}^{\prime}\right) 
 		\delta m^{(0)} \left(\mathbf{x}^{\prime}\right) \mathbf{u}^{(b)}\left(\mathbf{x}^{\prime}\right)  \\
 		&-\sum_{p=1}^{21} \int d \mathbf{x}^{\prime} \nabla_{\mathbf{x}^{\prime}} \mathbf{G}^{(b)}\left(\mathbf{r}, \mathbf{x}^{\prime}\right)
 		\cdot \left[ \mathbf{B}^{(p)} \delta m^{(p)} \left(\mathbf{x}^{\prime}\right) : \nabla_{\mathbf{x}^{\prime}} \mathbf{u}^{(b)}\left(\mathbf{x}^{\prime}\right)\right].	
 	\end{aligned}
 \end{equation}
 From equation \eqref{duB},	the Fr{\'e}chet derivative operators corresponding to different parameters can be defined as
	\begin{equation}\label{F}
		\begin{aligned}
			&\left[\mathcal{F}^{(0)} \delta \mathbf{m}^{(0)} \right] \left(\mathbf{r}\right) = 
			\omega^{2} \int d \mathbf{x}^{\prime} \mathbf{G}^{(b)}\left(\mathbf{r},\mathbf{x}^{\prime}\right)  \mathbf{u}^{(b)}\left(\mathbf{x}^{\prime}\right) \delta m^{(0)} \left(\mathbf{x}^{\prime}\right), \\
			&\left[\mathcal{F}^{(p)} \delta \mathbf{m}^{(p)} \right] \left(\mathbf{r}\right) = 
			-\int d \mathbf{x}^{\prime} \nabla_{\mathbf{x}^{\prime}} \mathbf{G}^{(b)}\left(\mathbf{r}, \mathbf{x}^{\prime}\right)
			\cdot \left[ \mathbf{B}^{(p)} : \nabla_{\mathbf{x}^{\prime}} \mathbf{u}^{(b)}\left(\mathbf{x}^{\prime}\right) \delta m^{(p)} \left(\mathbf{x}^{\prime}\right)\right].			
		\end{aligned}
	\end{equation}
	Here, $ \mathcal{F}^{(0)} $ is related to density perturbation and $ \mathcal{F}^{(p)} $ is related to elastic parameters perturbation. Inserting \eqref{F} into \eqref{duB} yields
	\begin{equation}\label{duF}
		\delta \mathbf{u}\left(\mathbf{r}\right) \approx 
		\left[\mathcal{F}^{(0)} \delta \mathbf{m}^{(0)} \right] \left(\mathbf{r}\right)
		+\sum_{p=1}^{21} \left[\mathcal{F}^{(p)} \delta \mathbf{m}^{(p)} \right] \left(\mathbf{r}\right).
	\end{equation}
	Equation \eqref{duB} can be further rewritten as
	\begin{equation}\label{duF2}
		\delta \mathbf{u} \approx \mathcal{F} \delta \mathbf{m},
	\end{equation}
	where $ \mathcal{F} = [\mathcal{F}^{(0)} ,\mathcal{F}^{(1)} ,...,\mathcal{F}^{(21)} ] $ and $ \delta \mathbf{m} = [\delta \mathbf{m}^{(0)} ,\delta \mathbf{m}^{(1)} ,...,\delta \mathbf{m}^{(21)} ]^{T} $.
	
	In most seismic applications, equation \eqref{duF2} is ill-posed due to the Fr{\'e}chet derivative operator being many-to-one rather than one-to-one. As a result, the solution of equation \eqref{duF2} is not unique. To obtain an appropriate solution to \eqref{duF2}, the generalized Tikhonov method is employed to ensure the stability of the solution. In the generalized Tikhonov method, a regularized solution to \eqref{duF2} is found by minimizing the objective function, as described in \citep{menke2012geophysical}:
	\begin{equation}\label{obj}
		\mathcal{E} \left(\delta \mathbf{m}\right)=\left\|\delta \mathbf{u} - \mathcal{F} \delta \mathbf{m}\right\|_{2}^{2}+ \lambda \left\|\delta \mathbf{m}\right\|_{2}^{2},
	\end{equation}
	in which $ \left\| \cdot \right\|_{2}^{2} $ is the $ L_{2} $ norm and $ \lambda $ is the regularization parameter. This paper determines the regularization parameter using a self-adaptive cooling scheme  \citep{jakobsen2015full}. An initial value for the regularization parameter is set at the beginning of the iteration process and is progressively reduced as the iterations proceed. To find the minimizer of $ \delta \mathbf{m} $ of \eqref{obj}, the following normal equation is utilized: 
	\begin{equation}\label{norm}
		(\mathcal{H} + \lambda \mathcal{I}) \delta \mathbf{m}= \mathcal{F}^{\dagger} \delta \mathbf{u},	
	\end{equation}
	where $ \mathcal{H} = \mathcal{F}^{\dagger} \mathcal{F} $ is the corresponding approximate Hessian operator, $ \mathcal{I} $ is the identity operator, and $ \mathcal{F}^{\dagger} $ is the adjoint of the Fr{\'e}chet operator. By utilizing the definition of adjoint \citep{tarantola2005inverse,claerbout2008image}, we have 
		\begin{equation}\label{ad}
			\begin{aligned} 
				& \left[(\mathcal{F}^{(0)})^{\dagger} \delta \mathbf{u}\right]\left(\mathbf{x}\right)
				= \left[\omega^{2} \int d \mathbf{r}  \mathbf{G}^{(b)}\left(\mathbf{x},\mathbf{r}\right) \delta \mathbf{u}^{*}\left(\mathbf{r}\right) \mathbf{u}^{(b)}\left(\mathbf{x}\right) \right]^{*},  \\	
				& \left[(\mathcal{F}^{(p)})^{\dagger} \delta \mathbf{u} \right]\left(\mathbf{x}\right)
				= \left[-\nabla_{\mathbf{x}}\int d \mathbf{r} \mathbf{G}^{(b)}\left(\mathbf{x},\mathbf{r}\right)  \delta \mathbf{u}^{*}\left(\mathbf{r}\right) 
				\cdot \mathbf{B}^{(p)} :\nabla_{\mathbf{x}} \mathbf{u}^{(b)} \left(\mathbf{x}\right)\right]^{*}, \\
			\end{aligned}
		\end{equation} 
where $ \left[\cdot\right]^{*} $ denotes complex conjugation. More details on the derivation of the adjoint operator can be found in Appendix~\ref{op_ad}. 

\subsection{Matrix-free formulations of the Fr{\'e}chet and adjoint operator}

In the normal equation \eqref{norm}, the model perturbation $ \delta \mathbf{m} $ can be solved by inverting $ \mathcal{H} + \lambda \mathcal{I}$. However, direct construct and inversion of $ \mathcal{H} + \lambda \mathcal{I}$ become computationally expensive and memory-intensive when dealing with realistic-scale problems. Instead, we solve the normal equation \eqref{norm} via a Krylov subspace method in conjunction with the matrix-free implementation of the Fr{\'e}chet and adjoint operator. By considering $ \mathbf{u}^{(b)} \delta m^{(0)} $ and $ \mathbf{B}^{(p)} : \nabla_{\mathbf{x}^{\prime}} \mathbf{u}^{(b)} \delta m^{(p)} $ in equation \eqref{F} as two virtual sources, denoted as $ \mathbf{Y} $ and $ \mathbf{M} $, we obtain
\begin{equation}\label{YM}
\begin{aligned}
\left[\mathcal{F}^{(0)} \delta \mathbf{m}^{(0)} \right] \left(\mathbf{r}\right) & = \omega^{2} \int d \mathbf{x}^{\prime} \mathbf{G}^{(b)}\left(\mathbf{r},\mathbf{x}^{\prime}\right)  \mathbf{Y} \left(\mathbf{x}^{\prime}\right), \\
\left[\mathcal{F}^{(p)} \delta \mathbf{m}^{(p)} \right] \left(\mathbf{r}\right) 
& = -\int d \mathbf{x}^{\prime} \nabla_{\mathbf{x}^{\prime}} \mathbf{G}^{(b)}\left(\mathbf{r}, \mathbf{x}^{\prime}\right)
\cdot \mathbf{M}\left(\mathbf{x}^{\prime}\right). \\
\end{aligned}
\end{equation}
Here, $ \mathbf{Y} $ is a vector source, while $ \mathbf{M} $ is a second rank moment tensor source. Drawing inspiration from the works of \citet{hesford2006frequency,hesford2010fast} and \citet{jakobsen2023seismic}, the physical interpretations of the two sub-equations in \eqref{YM} can be interpreted as the observed displacement fields from the background media due to sources $ \mathbf{Y} $ and $ \mathbf{M} $. 
Consequently, equation \eqref{YM} can be expressed as:
\begin{equation}\label{F0}
			\left[\mathcal{F}^{(0)} \delta \mathbf{m}^{(0)} \right] \left(\mathbf{r}\right) = \omega^{2} \mathbf{u}^{(Y)} \left(\mathbf{r}\right), \qquad
			\left[\mathcal{F}^{(p)} \delta \mathbf{m}^{(p)} \right] \left(\mathbf{r}\right) = -\mathbf{u}^{(M)}\left(\mathbf{r}\right), 
\end{equation}
where the observed displacement fields $ \mathbf{u}^{(Y)} \left(\mathbf{r}\right) $ and $ \mathbf{u}^{(M)}\left(\mathbf{r}\right) $ can be solved from 
\begin{equation}\label{uYr}
	\begin{aligned} 
		&\begin{aligned} 
			\mathbf{u}^{(Y)} \left(\mathbf{r}\right) = 
			& \mathbf{u}^{(Y,0)}\left(\mathbf{r}\right)+ \omega^{2} \int d \mathbf{x}^{\prime} \mathbf{G}^{(0)}\left(\mathbf{r}- \mathbf{x}^{\prime}\right) \delta \rho^{(b)}\left(\mathbf{x}^{\prime}\right) \mathbf{u}^{(Y)} \left(\mathbf{x}^{\prime}\right) \\
			& -\int d \mathbf{x}^{\prime} \nabla_{\mathbf{x}^{\prime}} \mathbf{G}^{(0)}\left(\mathbf{r}- \mathbf{x}^{\prime}\right)
			\cdot \delta \mathbf{C}^{(b)}\left(\mathbf{x}^{\prime}\right):\nabla_{\mathbf{x}^{\prime}} \mathbf{u}^{(Y)} \left(\mathbf{x}^{\prime}\right),
		\end{aligned}  \\
		&\begin{aligned} 
			\mathbf{u}^{(M)} \left(\mathbf{r}\right) = 
			& \mathbf{u}^{(M,0)}\left(\mathbf{r}\right)+ \omega^{2} \int d \mathbf{x}^{\prime} \mathbf{G}^{(0)}\left(\mathbf{r}- \mathbf{x}^{\prime}\right) \delta \rho^{(b)}\left(\mathbf{x}^{\prime}\right) \mathbf{u}^{(M)} \left(\mathbf{x}^{\prime}\right) \\
			& -\int d \mathbf{x}^{\prime} \nabla_{\mathbf{x}^{\prime}} \mathbf{G}^{(0)}\left(\mathbf{r}- \mathbf{x}^{\prime}\right)
			\cdot \delta \mathbf{C}^{(b)}\left(\mathbf{x}^{\prime}\right):\nabla_{\mathbf{x}^{\prime}} \mathbf{u}^{(M)} \left(\mathbf{x}^{\prime}\right),
		\end{aligned}  
	\end{aligned} 
\end{equation}
where 
\begin{equation}\label{key}
	\delta \rho^{(b)} = \rho^{(b)} - \rho^{(0)}, \qquad \delta \mathbf{C}^{(b)} = \mathbf{C}^{(b)} - \mathbf{C}^{(0)}.
\end{equation}
To solve equation \eqref{uYr} involves knowledge of the displacement fields $ \mathbf{u}^{(Y)}\left(\mathbf{x}^{\prime}\right) $ and $ \mathbf{u}^{(M)}\left(\mathbf{x}^{\prime}\right) $  within the heterogeneous background media, which we need to first compute by solving the forward-scattering equation
\begin{equation}\label{uY}
	\begin{aligned} 
		&\begin{aligned} 
			\mathbf{u}^{(Y)} \left(\mathbf{x}\right) = 
			& \mathbf{u}^{(Y,0)}\left(\mathbf{x}\right)+ \omega^{2} \int d \mathbf{x}^{\prime} \mathbf{G}^{(0)}\left(\mathbf{x}- \mathbf{x}^{\prime}\right) \delta \rho^{(b)}\left(\mathbf{x}^{\prime}\right) \mathbf{u}^{(Y)} \left(\mathbf{x}^{\prime}\right) \\
			& -\int d \mathbf{x}^{\prime} \nabla_{\mathbf{x}^{\prime}} \mathbf{G}^{(0)}\left(\mathbf{x}- \mathbf{x}^{\prime}\right)
			\cdot \delta \mathbf{C}^{(b)}\left(\mathbf{x}^{\prime}\right):\nabla_{\mathbf{x}^{\prime}} \mathbf{u}^{(Y)} \left(\mathbf{x}^{\prime}\right),
		\end{aligned}  \\
		&\begin{aligned} 
			\mathbf{u}^{(M)} \left(\mathbf{x}\right) = 
			& \mathbf{u}^{(M,0)}\left(\mathbf{x}\right)+ \omega^{2} \int d \mathbf{x}^{\prime} \mathbf{G}^{(0)}\left(\mathbf{x}- \mathbf{x}^{\prime}\right) \delta \rho^{(b)}\left(\mathbf{x}^{\prime}\right) \mathbf{u}^{(M)} \left(\mathbf{x}^{\prime}\right) \\
			& -\int d \mathbf{x}^{\prime} \nabla_{\mathbf{x}^{\prime}} \mathbf{G}^{(0)}\left(\mathbf{x}- \mathbf{x}^{\prime}\right)
			\cdot \delta \mathbf{C}^{(b)}\left(\mathbf{x}^{\prime}\right):\nabla_{\mathbf{x}^{\prime}} \mathbf{u}^{(M)} \left(\mathbf{x}^{\prime}\right),
		\end{aligned}  \\
	\end{aligned} 
\end{equation}
where 
\begin{equation}\label{uM0}
	\begin{aligned} 
		&\mathbf{u}^{(Y,0)}\left(\mathbf{x}\right) 
		= \int d \mathbf{x}^{\prime} \mathbf{G}^{(0)}\left(\mathbf{x}-\mathbf{x}^{\prime}\right)  \mathbf{Y} \left(\mathbf{x}^{\prime}\right), \\
		&\mathbf{u}^{(M,0)}\left(\mathbf{x}\right) 
		= \int d \mathbf{x}^{\prime} \nabla_{\mathbf{x}^{\prime}} \mathbf{G}^{(0)}\left(\mathbf{x}-\mathbf{x}^{\prime}\right)  \mathbf{M} \left(\mathbf{x}^{\prime}\right) \\
	\end{aligned}  
\end{equation}
are the reference displacement fields.

In the same way, following the physical interpretations of the sub-equations in \eqref{ad}, the adjoint operators can be rewritten as 
\begin{equation}\label{ad1}
	\begin{aligned} 
		& \left[(\mathcal{F}^{(0)})^{\dagger} \delta \mathbf{u}\right]\left(\mathbf{x}\right)
		= \left[\omega^{2} \mathbf{u}^{(a)}\left(\mathbf{x}\right) \mathbf{u}^{(b)}\left(\mathbf{x}\right) \right]^{*},  \\	
		& \left[(\mathcal{F}^{(p)})^{\dagger} \delta \mathbf{u} \right]\left(\mathbf{x}\right)
		= \left[-\nabla_{\mathbf{x}} \mathbf{u}^{(a)} \left(\mathbf{x}\right)
		\cdot \mathbf{B}^{(p)} :\nabla_{\mathbf{x}} \mathbf{u}^{(b)} \left(\mathbf{x}\right)\right]^{*}, \\
	\end{aligned}
\end{equation} 
where 
\begin{equation}\label{ua}
	\mathbf{u}^{(a)} \left(\mathbf{x}\right)= \int d \mathbf{r}  \mathbf{G}^{(b)}\left(\mathbf{x},\mathbf{r}\right) \delta \mathbf{u}^{*}\left(\mathbf{r}\right).
\end{equation} 
Here, $ \mathbf{u}^{(a)} $ is the back-propagating displacement field within the heterogeneous background media from source $ \delta \mathbf{u}^{*} $, which can be solved from 
\begin{equation}\label{ua}
	\begin{aligned} 
		&\begin{aligned} 
			\mathbf{u}^{(a)} \left(\mathbf{x}\right) = 
			& \mathbf{u}^{(a,0)}\left(\mathbf{x}\right)+ \omega^{2} \int d \mathbf{x}^{\prime} \mathbf{G}^{(0)}\left(\mathbf{x}- \mathbf{x}^{\prime}\right) \delta \rho^{(b)}\left(\mathbf{x}^{\prime}\right) \mathbf{u}^{(a)} \left(\mathbf{x}^{\prime}\right) \\
			& -\int d \mathbf{x}^{\prime} \nabla_{\mathbf{x}^{\prime}} \mathbf{G}^{(0)}\left(\mathbf{x}- \mathbf{x}^{\prime}\right)
			\cdot \delta \mathbf{C}^{(b)}\left(\mathbf{x}^{\prime}\right):\nabla_{\mathbf{x}^{\prime}} \mathbf{u}^{(a)} \left(\mathbf{x}^{\prime}\right),
		\end{aligned}  \\
	\end{aligned} 
\end{equation}
where 
\begin{equation}\label{ua0}
	\begin{aligned} 
		&\mathbf{u}^{(a,0)}\left(\mathbf{x}\right) 
		= \int d \mathbf{r} \mathbf{G}^{(0)}\left(\mathbf{x}-\mathbf{r}\right)  \delta \mathbf{u}^{*}\left(\mathbf{r}\right). \\
	\end{aligned}  
\end{equation}
Solving equations \eqref{uY} and \eqref{ua} are forward modeling problems that can be effectively solved utilizing the methods mentioned in the preceding section. The integrals involving Green's function and its derivative in all the above equations can be treated as convolutions and can be efficiently and accurately computed using the fast Fourier transform. This eliminates the need to store the matrix of the homogeneous Green's function and compute the Green's function for the heterogeneous background medium. The current formulations of the Fr{\'e}chet derivative operator \eqref{F0} and its adjoint operator \eqref{ad1} only involve vectors: $\mathbf{u}^{(Y)}$, $\mathbf{u}^{(M)}$, $\mathbf{u}^{(a)}$, $\mathbf{u}^{(b)}$, $\nabla_{\mathbf{x}^{\prime}} \mathbf{u}^{(a)}$, and $\nabla_{\mathbf{x}^{\prime}} \mathbf{u}^{(b)}$, making our method matrix-free.

				\subsection{Abbreviated subscript notation for implementation}
				
				The components of the elastic stiffness tensor exhibit symmetries ($ C_{ijkl}=C_{jikl}=C_{ijlk}=C_{jilk} $); thus, employing the abbreviated subscript notation \citep{auld1973acoustic}, the tensor can be compactly represented as a 6$\times$6 matrix:
				\begin{equation}
				\mathbf{C}=\left[\begin{array}{cccccc}
				C_{11} & C_{12} & C_{13} & C_{14} & C_{15} & C_{16}\\
				C_{12} & C_{22} & C_{23} & C_{24} & C_{25} & C_{26}\\
				C_{13} & C_{23} & C_{33} & C_{34} & C_{35} & C_{36}\\
				C_{14} & C_{24} & C_{34} & C_{44} & C_{45} & C_{46}\\
				C_{15} & C_{25} & C_{35} & C_{45} & C_{55} & C_{56}\\
				C_{16} & C_{26} & C_{36} & C_{46} & C_{56} & C_{66}
				\end{array}\right].
				\end{equation}
				In the same way, by using the abbreviated subscript notation, 
				the tensors representing the derivatives of the Green's function $ \nabla \mathbf{G}^{(0)} $ and the displacement gradient $ \nabla \mathbf{u} $ can be  represented as:
				\begin{equation}
				\nabla_{\mathbf{x}} \mathbf{G}^{(0)} =
				\left[\begin{array}{cccccc}
				\frac{\partial G^{(0)}_{11}}{\partial x_1} & \frac{\partial G^{(0)}_{12}}{\partial x_2} &\frac{\partial G^{(0)}_{13}}{\partial x_3} &\frac{\partial G^{(0)}_{12}}{\partial x_3}+\frac{\partial G^{(0)}_{13}}{\partial x_2} &\frac{\partial G^{(0)}_{11}}{\partial x_3}+\frac{\partial G^{(0)}_{13}}{\partial x_1} & \frac{\partial G^{(0)}_{11}}{\partial x_2}+\frac{\partial G^{(0)}_{12}}{\partial x_1} \\
				\frac{\partial G^{(0)}_{21}}{\partial x_1} & \frac{\partial G^{(0)}_{22}}{\partial x_2} &\frac{\partial G^{(0)}_{23}}{\partial x_3} &\frac{\partial G^{(0)}_{22}}{\partial x_3}+\frac{\partial G^{(0)}_{23}}{\partial x_2} &\frac{\partial G^{(0)}_{21}}{\partial x_3}+\frac{\partial G^{(0)}_{23}}{\partial x_1} & \frac{\partial G^{(0)}_{21}}{\partial x_2}+\frac{\partial G^{(0)}_{22}}{\partial x_1} \\
				\frac{\partial G^{(0)}_{31}}{\partial x_1} & \frac{\partial G^{(0)}_{32}}{\partial x_2} &\frac{\partial G^{(0)}_{33}}{\partial x_3} &\frac{\partial G^{(0)}_{32}}{\partial x_3}+\frac{\partial G^{(0)}_{33}}{\partial x_2} &\frac{\partial G^{(0)}_{31}}{\partial x_3}+\frac{\partial G^{(0)}_{33}}{\partial x_1} & \frac{\partial G^{(0)}_{31}}{\partial x_2}+\frac{\partial G^{(0)}_{32}}{\partial x_1} \\
				\end{array}\right],
				\end{equation}
				and
				\begin{equation}
				\nabla_{\mathbf{x}} \mathbf{u}
				=\left[\begin{array}{cccccc}
				\frac{\partial u_1}{\partial x_1} &
				\frac{\partial u_2}{\partial x_2} &
				\frac{\partial u_3}{\partial x_3} &
				\frac{\partial u_2}{\partial x_3}+\frac{\partial u_3}{\partial x_2} &
				\frac{\partial u_1}{\partial x_3}+\frac{\partial u_3}{\partial x_1} &
				\frac{\partial u_1}{\partial x_2}+\frac{\partial u_2}{\partial x_1} 
				\end{array}\right]^{T},
				\end{equation}
				where $ T $ denotes the transpose. The matrix-free distorted Born iterative method proposed in this study has the theoretical capability to invert all 21 elastic parameters and density. However, in practical applications, inverting such a vast number of parameters simultaneously is not feasible due to the significant computation and storage resources required.	Therefore, we restrict our numerical tests to the transversely isotropic (VTI) media with a vertical symmetry axis with variable density. The VTI media are generally described by five independent elastic parameters: $ C_{11} $, $ C_{33} $, $ C_{55} $, $ C_{66} $, and $ C_{13} $ \citep{carcione2014wave}. 
				The simplified elasticity tensor corresponding to VTI media is presented as follows:
								\begin{equation}
								\mathbf{C}^{V T I}=\left[\begin{array}{cccccc}
								C_{11} & C_{12} & C_{13} & 0 & 0 & 0 \\
								C_{12} & C_{11} & C_{13} & 0 & 0 & 0 \\
								C_{13} & C_{13} & C_{33} & 0 & 0 & 0 \\
								0 & 0 & 0 & C_{55} & 0 & 0 \\
								0 & 0 & 0 & 0 & C_{55} & 0 \\
								0 & 0 & 0 & 0 & 0 & C_{66}
								\end{array}\right], \quad C_{12} = C_{11}-2 C_{66}.
								\end{equation} 
				Despite our focus on inverting only five independent elastic parameters and density, as opposed to the full set of 22 parameters, this remains an immensely challenging task due to the substantial computational cost and crosstalk issues inherent to multi-parameter inversion. In the abbreviated subscript notation, the tensor $ \mathbf{B} $ in equation \eqref{dC} has also been reformulated as a $ 6 \times 6 $ constant matrix. The $ \mathbf{B} $ matrices for different elastic parameters of the VTI media are shown in Appendix~\ref{B}.			

				\section*{Numerical examples}
				\subsection{2D VTI Reservoir model}

				We first tested the proposed method on a simple 2D VTI Reservoir model. This model is described by four elastic parameters ($ C_{11} $, $ C_{33} $, $ C_{55} $, $ C_{13} $) and density $ \rho $, as shown in Figure~\ref{fig:true_res_2D4}. The size of this model is 3600 m in the horizontal dimension and 900 m in the depth dimension. It has been discretized into 180 $\times$ 45 grid blocks for numerical computation. There are 90 receivers and 45 sources uniformly distributed at the top of this model. A Ricker wavelet with a central frequency of 10 Hz has been used to generate the incident wave. In this example, we employ the sequential frequency inversion scheme, in which frequencies are inverted individually from the lowest to the highest. The nine frequencies used for this example are 3 Hz, 5 Hz, 7 Hz, 9 Hz, 11 Hz, 13 Hz, 15 Hz, 17 Hz, and 19 Hz. We use the GMRES (generalized minimum residual) method with the fast Fourier transform acceleration to solve the integral equation \eqref{u_2} for the calculation of displacement fields. To accurately quantify the disparity between the calculated data $ \mathbf{d}_{cal} $ and the observed data $ \mathbf{d}_{obs} $ as well as the difference between the true model $ \mathbf{m}_{true} $ and the updated model $ \mathbf{m} $ during the inversion process, we define the normalized data error $ \epsilon_{d} $ and normalized model difference $ \epsilon_{m} $ as
				\begin{equation}
				\epsilon_{d} = \frac{\left\|\mathbf{d}_{cal} - \mathbf{d}_{obs}\right\| }{\left\|\mathbf{d}_{obs}\right\|},
				\end{equation}
				\begin{equation}
				\epsilon_{m} = \frac{\left\|\mathbf{m} - \mathbf{m}_{true}\right\|}{\left\|\mathbf{m}_{true}\right\|}.
				\end{equation}
				The stopping criteria of model update for each frequency are either (1) data error $ \epsilon_{d} $ less than 0.001 or (2) the number of iterations reaches 10.  We created the initial models, as shown in Figure~\ref{fig:ini_res_2D4}, by applying a Gaussian smoothing filter with a standard deviation of 15 to the true models.

			    The inverted results are shown in Figure~\ref{fig:inv_res_2D4}. All of the model parameters are well recovered. However, we can see that the inverted resolution of $ C_{13} $ and $ \rho $ in Figure~\ref{fig:inv_res_2D4} is not as good as the other parameters. This is because the physical parameters are coupled with each other, and the perturbation of the elastic parameters is much larger than the perturbation of the density. In this case, the strong mapping from other parameters greatly influences $ C_{13} $ and $ \rho $ during the inversion process. This can be reduced through different parameterization techniques \citep{operto2013guided,prieux2013multiparameter}, which we will try in our further	work. Figure~\ref{fig:conv_res_2D3} shows the convergence performance of our method. The upper and middle plots in Figure~\ref{fig:conv_res_2D3} show the normalized data error and model difference change with iteration. The lower plot in Figure~\ref{fig:conv_res_2D3} gives the related frequency for each iteration. From Figure~\ref{fig:conv_res_2D3}, we can see that the data error and model difference decrease with iteration during the inversion process at each frequency. All these results illustrate that our method has been successfully implemented in the full waveform inversion for the elastic anisotropic media.

\plot{true_res_2D4}{width=1\textwidth}
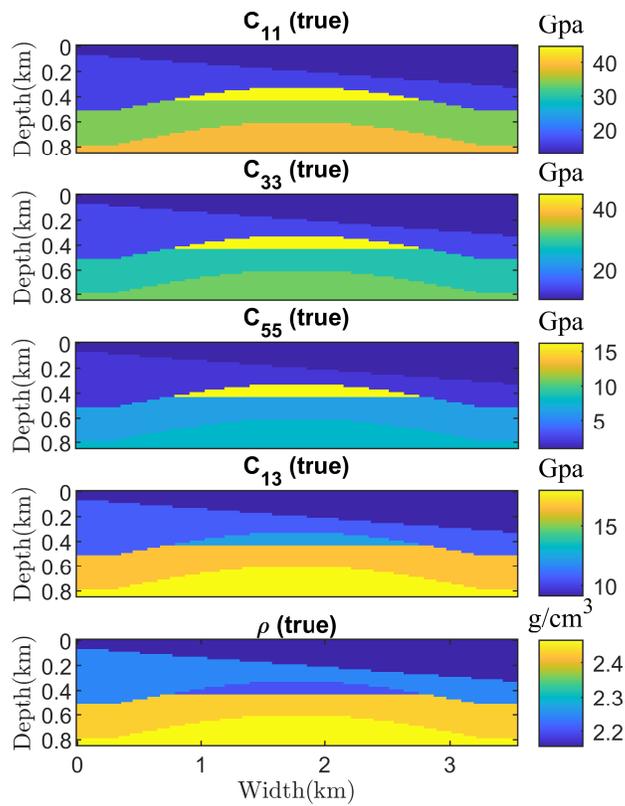
{2D reservoir model described by four elastic parameters and density.}

\plot{ini_res_2D4}{width=1\textwidth}
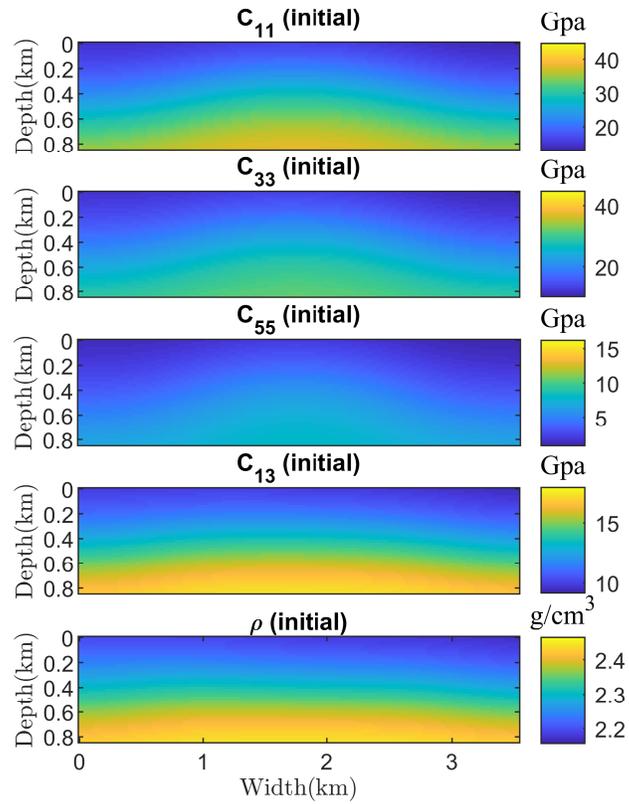
{Initial model with four elastic parameters and density obtained by smoothing the true model in Figure~\ref{fig:true_res_2D4}.}

\plot{inv_res_2D4}{width=1\textwidth}
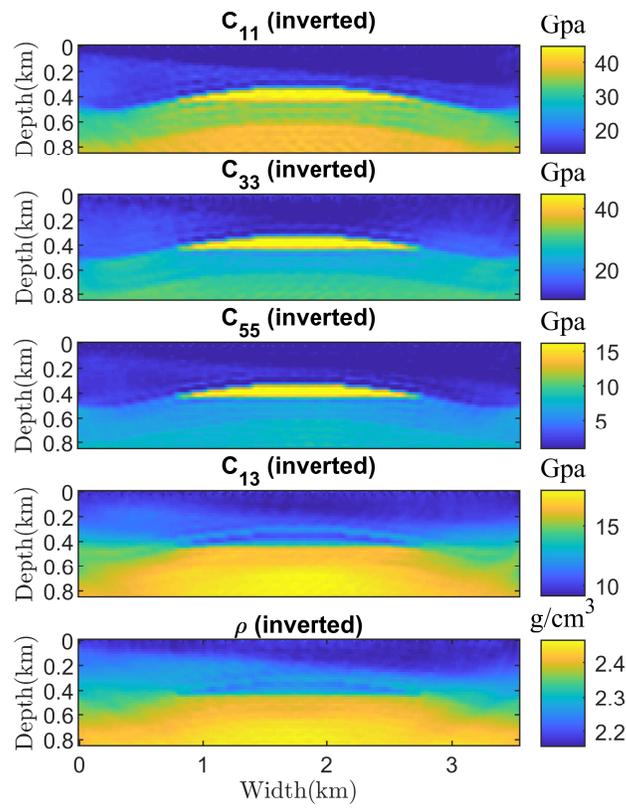
{Inverted results of the 2D reservoir model.}

\plot{conv_res_2D3}{width=0.9\textwidth}
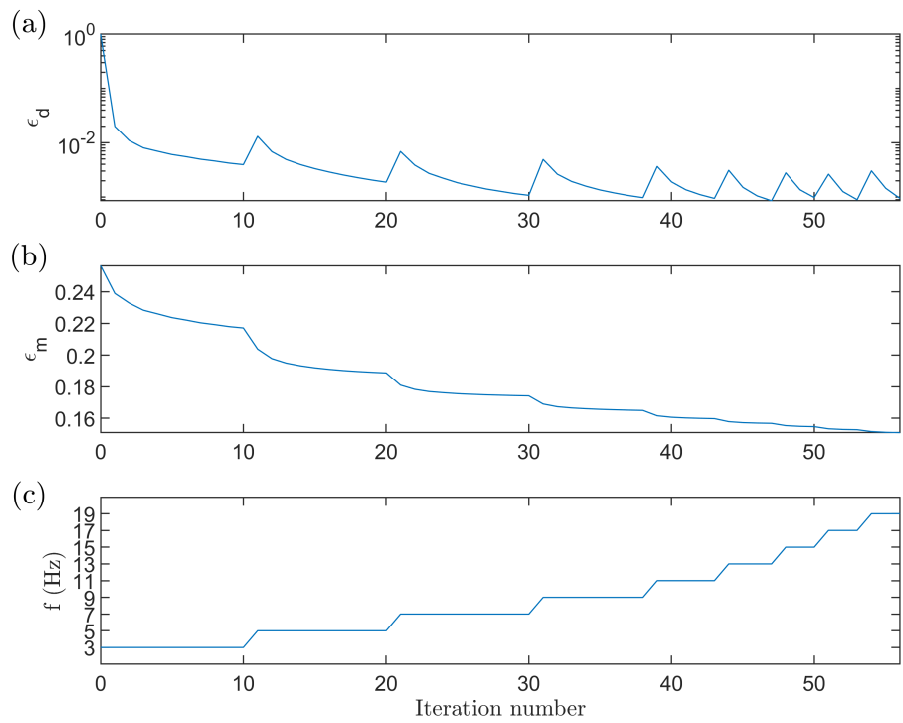
{Convergence diagrams of the 2D reservoir model test: (a) normalized data difference $ \epsilon_{d} $ versus the number of iterations, (b) normalized model error $ \epsilon_{m} $ versus the number of iterations, and (c) the frequency corresponding to each iteration.}

				\subsection{2D VTI Hess model}
				 We have used a resampled version of the 2D VTI Hess model to test our method on a more practical and complicated model (Figure~\ref{fig:true_Hess_2D3}). The size of this model is 3220 m in the horizontal dimension and 860 m in the depth dimension. It has been discretized into 161 $\times$ 43 grid blocks for numerical computation. There are 80 receivers and 40 sources uniformly distributed at the top of this model. A Ricker wavelet with a peak frequency of 10 Hz has been used to generate the incident wave. In this example, we employ eight frequency components for inversion: 3 Hz, 5 Hz, 7 Hz, 9 Hz, 11 Hz, 13 Hz, 15 Hz, and 17 Hz. All the displacement fields are computed by equation \eqref{u_2} with the fast-Fourier-transform-accelerated GMRES (generalized minimum residual) method. The stopping criteria of model update for each frequency are either (1) data residual $ \epsilon_{d} $ less than 0.001 or (2) the number of iterations reaches 10. We generate the initial models (Figure~\ref{fig:ini_Hess_2D3}) by filtering the true models with a Gaussian smoothing kernel with a standard deviation of 14.	
				 	
				The inverted stiffness parameters and density are shown in Figure~\ref{fig:inv_Hess_2D3}. All parameters have been accurately estimated, and the structure of this model has been well constructed. The fault is evident on the right side of the inverted model. Figure~\ref{fig:conv_Hess_2D} shows the normalized data error and model difference change with iterations for different frequencies. At each frequency, the data error and model difference decrease with iteration during the inversion process. These inversion results illustrate that our method can deal with a complicated model.
	
				\plot{true_Hess_2D3}{width=1\textwidth}
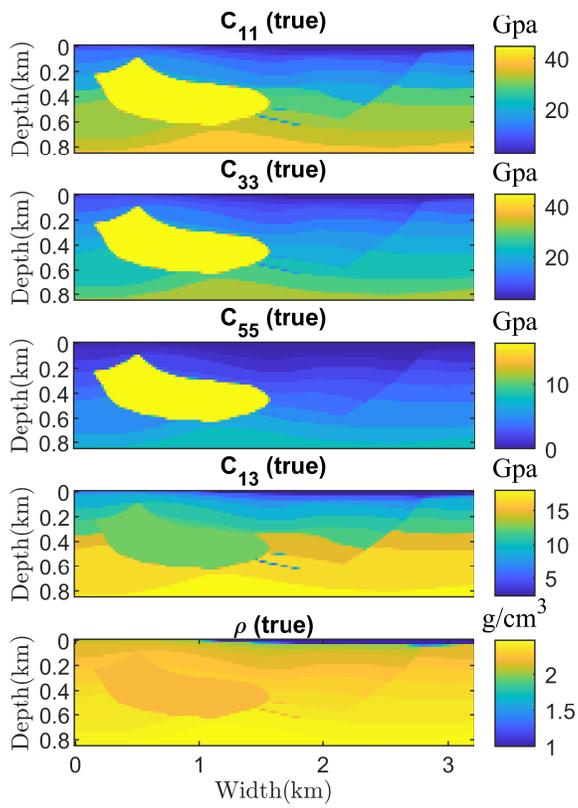
				{Resampled 2D VTI Hess model.}
				
				\plot{ini_Hess_2D3}{width=1\textwidth}
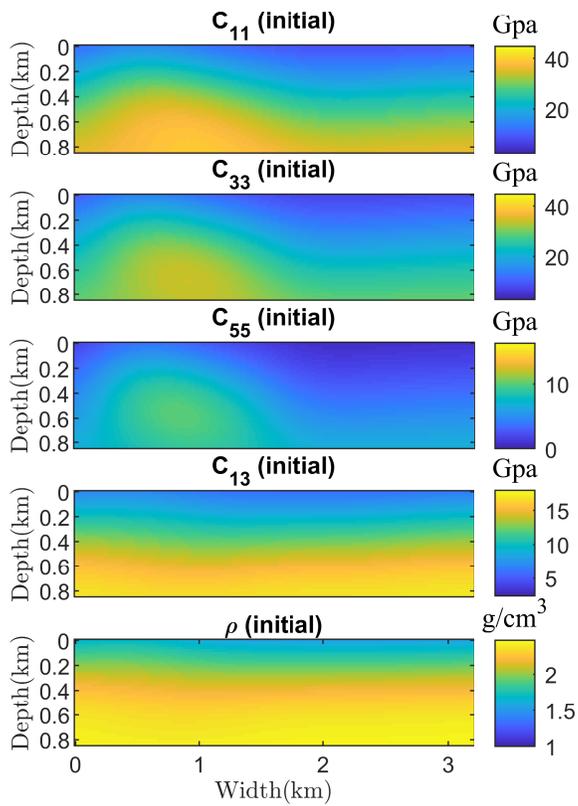
				{Initial model obtained by smoothing the true model in Figure~\ref{fig:true_Hess_2D3}.}
				
				\plot{inv_Hess_2D3}{width=1\textwidth}
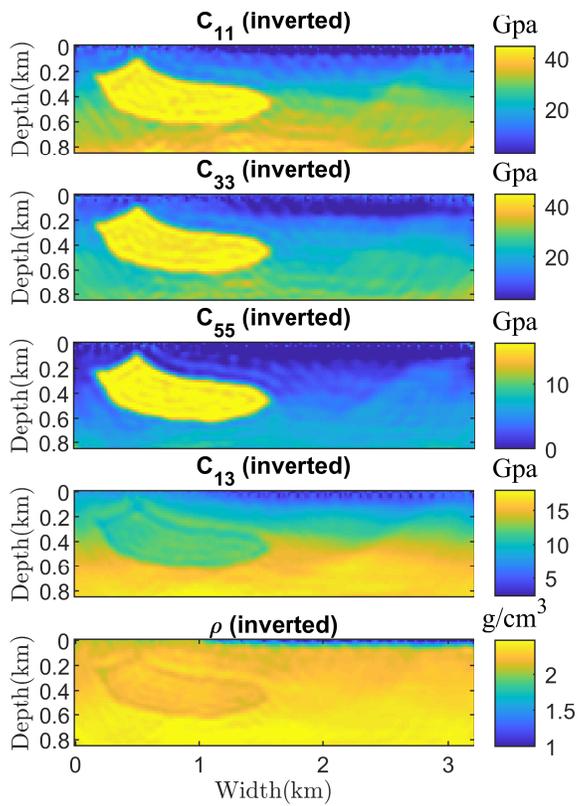
				{Inverted results of the resampled 2D VTI Hess model.}

				\plot{conv_Hess_2D}{width=0.9\textwidth}
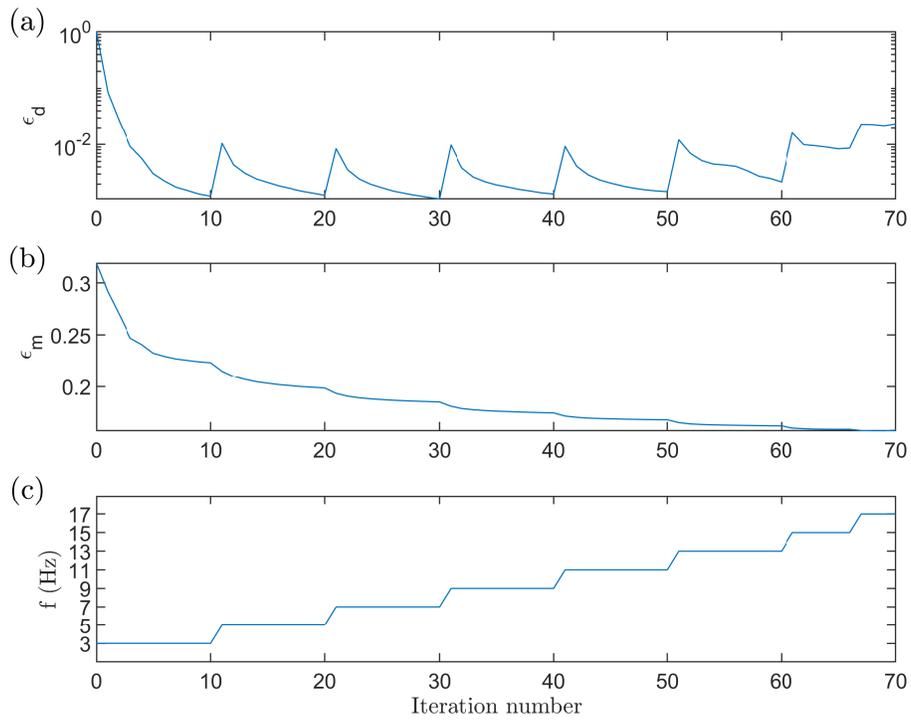
				{Convergence diagrams of the resampled 2D VTI Hess model test: (a) normalized data difference $ \epsilon_{d} $ versus the number of iterations, (b) normalized model error $ \epsilon_{m} $ versus the number of iterations, and (c) the frequency corresponding to each iteration.}

				\subsection{Modified 3D VTI Hess model}

Next, we extended the test of our method from 2D to 3D. In this test, a modified 3D Hess model is utilized, which expands upon the 2D Hess model in the y-direction. The 3D VTI model are described by six physical parameters ($ C_{11} $, $ C_{33} $, $ C_{55} $, $ C_{66} $, $ C_{13} $, and $\rho$), as shown in Figure~\ref{fig:true_C11_3Dvis,true_C33_3Dvis,true_C55_3Dvis,true_C66_3Dvis,true_C13_3Dvis,true_rho_3Dvis}. Here we set  $ C_{66} = C_{55} $, because the 2D model does not include $ C_{66} $. This modified 3D VTI Hess model's dimensions are 2000 m $ \times $1000 m $ \times $ 800 m. We discretize this model into 80$ \times $40$\times$32 grid blocks. Each grid block has a size 25m$ \times $25m$\times$25m. There are 100 sources and 400 receivers uniformly distributed on the top of this model, as shown in Figure~\ref{fig:SR7}. A Ricker wavelet with a dominant frequency of 10 Hz is used to generate the incident wavefield. The initial model (Figure~\ref{fig:ini_C11_3Dvis,ini_C33_3Dvis,ini_C55_3Dvis,ini_C66_3Dvis,ini_C13_3Dvis,ini_rho_3Dvis}) is a smooth version of the true model with a 3D Gaussian smoothing kernel with standard deviation 11. Seven frequencies, 3 Hz, 5 Hz, 7 Hz, 9 Hz, 11 Hz, 13 Hz, and 15 Hz, are used for this inversion. 

Figure~\ref{fig:inv_C11_3Dvis,inv_C33_3Dvis,inv_C55_3Dvis,inv_C66_3Dvis,inv_C13_3Dvis,inv_rho_3Dvis} shows the 3D inverted results. To show the inner structure of the 3D inverted results, we also show three slices in the y-direction for each inverted parameter in Figure~\ref{fig:inv_C11_3D7,inv_C33_3D7,inv_C55_3D7,inv_C66_3D7,inv_C13_3D7,inv_rho_3D7}. The parameters are accurately estimated, and the model's structure is well-defined. The inverted model clearly delineates the salt dome and the fault structure.
Figure~\ref{fig:conv_3D7} shows the normalized data error and model difference change during the inversion process at different frequencies. At each frequency, there is a consistent decrease in both data error and model difference throughout the inversion. All these figures show that the proposed method can also be used for the inversion of a 3D VTI model.

				\multiplot{2}{true_C11_3Dvis,true_C33_3Dvis,true_C55_3Dvis,true_C66_3Dvis,true_C13_3Dvis,true_rho_3Dvis}{width=0.45\textwidth}
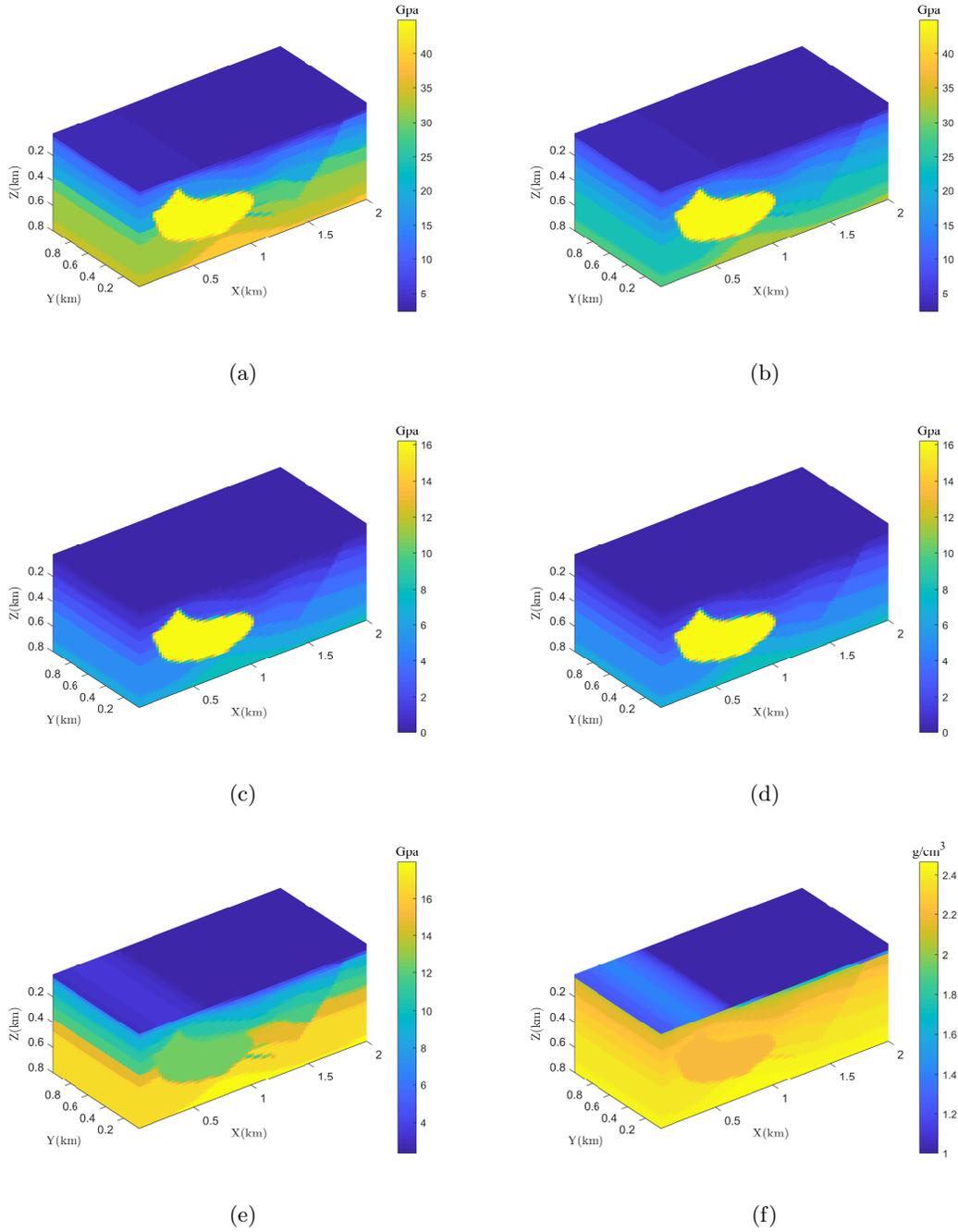
				{Modified 3D VTI Hess model:(a) $ C_{11} $, (b) $ C_{33} $, (c) $ C_{55} $, (d) $ C_{66} $, (e) $ C_{13} $, and (f) $ \rho $.}	
								\plot{SR7}{width=0.9\textwidth}
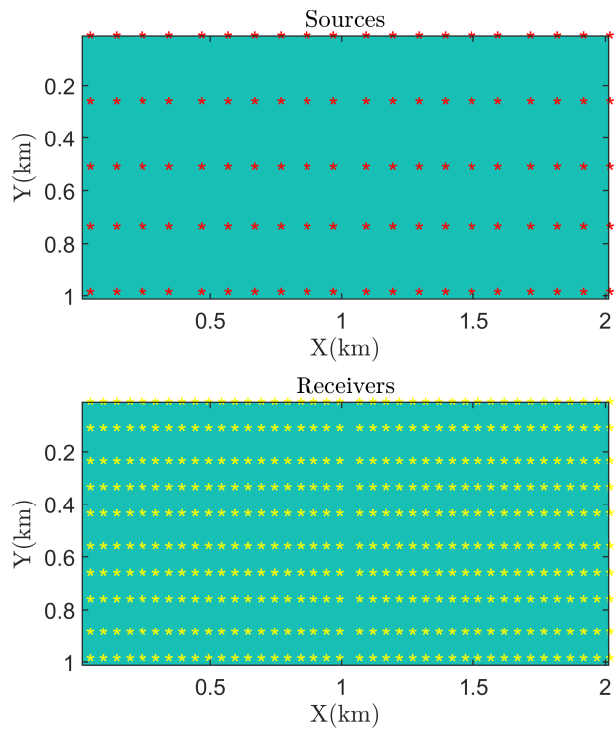
						{Distribution of sources and receivers on the top of the 3D model.}
\multiplot{2}{ini_C11_3Dvis,ini_C33_3Dvis,ini_C55_3Dvis,ini_C66_3Dvis,ini_C13_3Dvis,ini_rho_3Dvis}{width=0.45\textwidth}
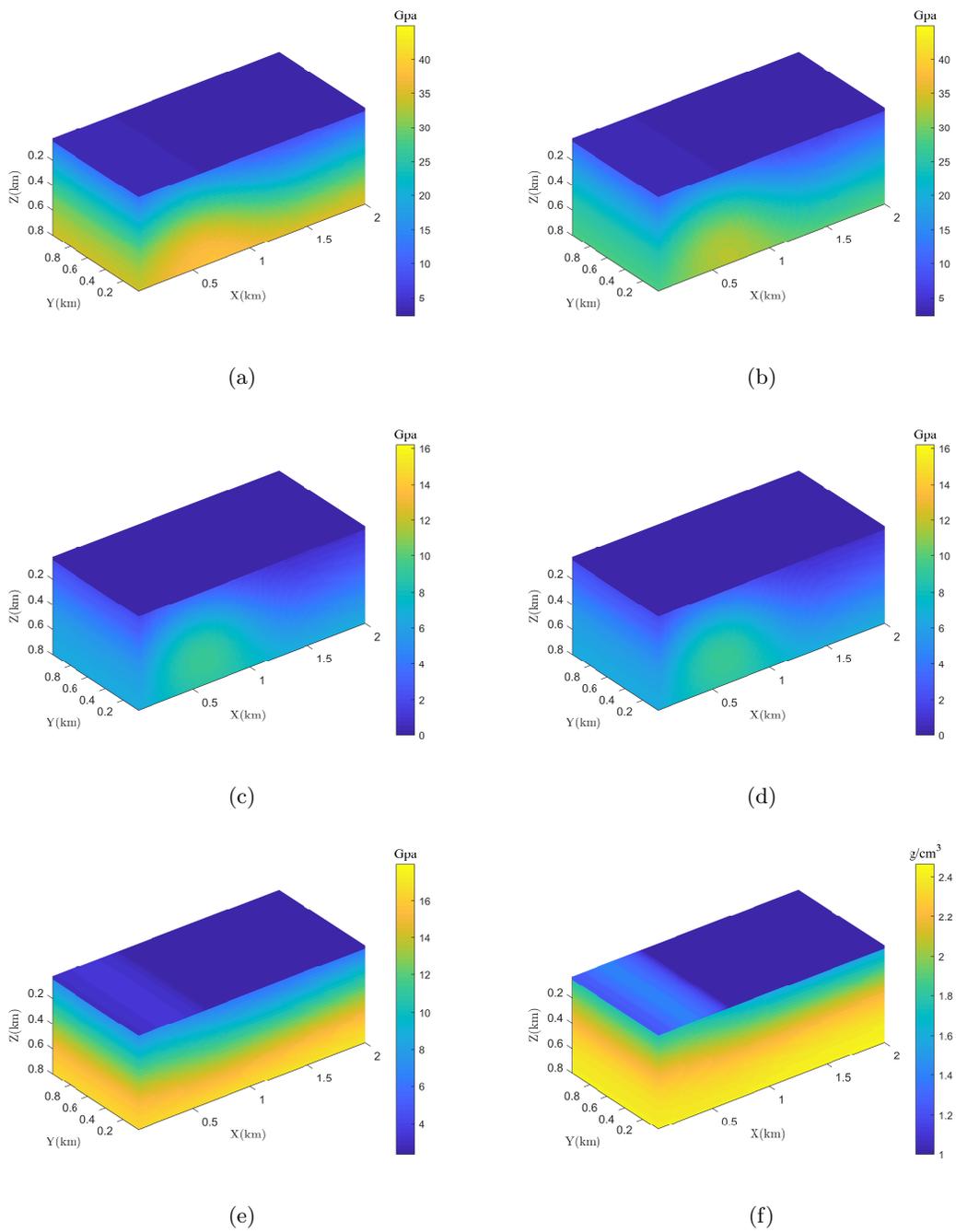
{3D initial model:(a) $ C_{11} $, (b) $ C_{33} $, (c) $ C_{55} $, (d) $ C_{66} $, (e) $ C_{13} $, and (f) $ \rho $.}
\multiplot{2}{inv_C11_3Dvis,inv_C33_3Dvis,inv_C55_3Dvis,inv_C66_3Dvis,inv_C13_3Dvis,inv_rho_3Dvis}{width=0.45\textwidth}
{3D inverted model:(a) $ C_{11} $, (b) $ C_{33} $, (c) $ C_{55} $, (d) $ C_{66} $, (e) $ C_{13} $, and (f) $ \rho $.}
\multiplot{2}{inv_C11_3D7,inv_C33_3D7,inv_C55_3D7,inv_C66_3D7,inv_C13_3D7,inv_rho_3D7}{width=0.45\textwidth}
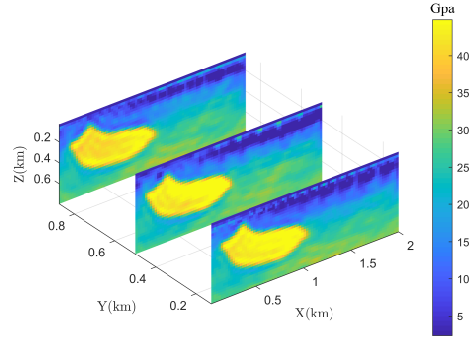
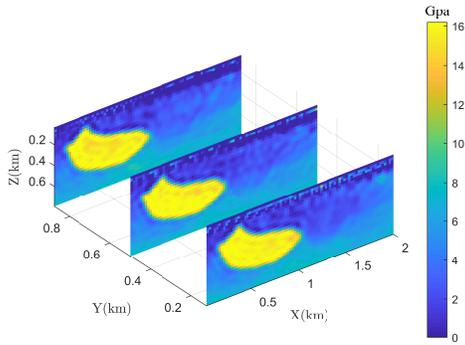
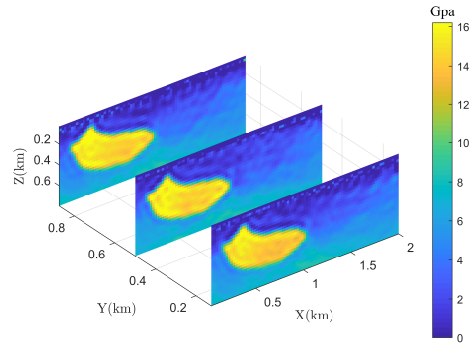
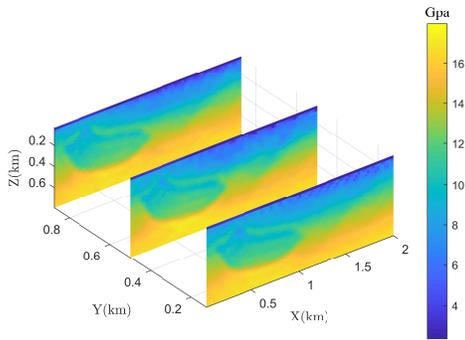
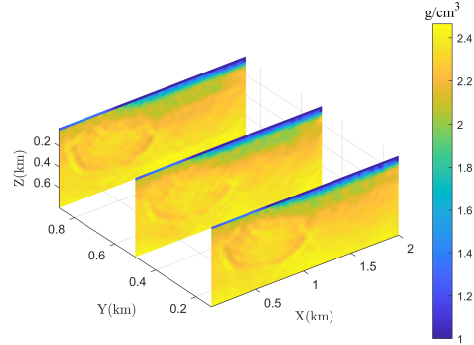
				{Slice visualization of the 3D inverted model:(a) $ C_{11} $, (b) $ C_{33} $, (c) $ C_{55} $, (d) $ C_{66} $, (e) $ C_{13} $, and (f) $ \rho $.}
	
				\plot{conv_3D7}{width=0.9\textwidth}
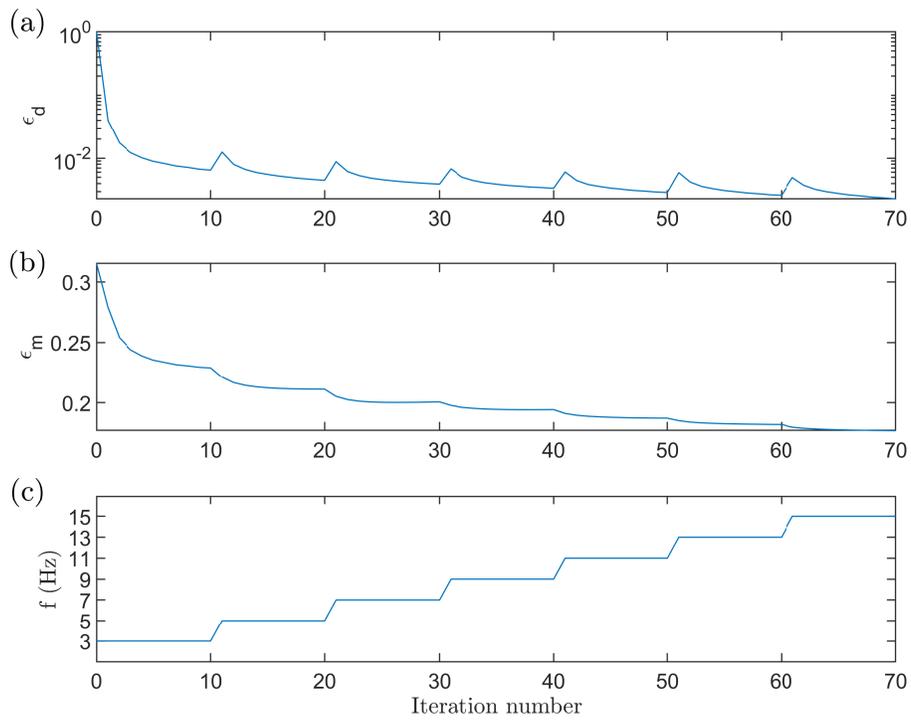
				{Convergence diagrams of the modified 3D VTI Hess model test: (a) normalized data difference $ \epsilon_{d} $ versus the number of iterations, (b) normalized model error $ \epsilon_{m} $ versus the number of iterations, and (c) the frequency corresponding to each iteration.} 
				
				\section{Concluding remarks}

				We have extended the application of the matrix-free distorted Born iterative method to multi-parameter full waveform inversion that can reconstruct elastic parameters and density simultaneously from frequency-domain waveform data. The key idea of our method is the matrix-free implementations of the Fr{\'e}chet derivatives of different parameters and their adjoint operators, which can significantly reduce the computational cost and memory demand.
				In multi-parameter inversion, the Hessian information is crucial for reducing the crosstalk effects among different parameters. The newly proposed algorithm effectively utilizes the Hessian information by incorporating Fr{\'e}chet derivatives and their adjoint operators, eliminating the need for forming and inverting the entire Hessian matrix. 
				The Fr{\'e}chet derivatives and their adjoint operators are formulated as vector operations according to the physical meaning of Green's function. In our formulations, applying the Fr{\'e}chet derivative and its adjoint does not require knowledge of Green's function in the heterogeneous background media. Instead, a set of virtual displacement fields is computed using an integral equation solver with fast Fourier transform acceleration. All these improvements make the distorted Born iterative method more practical for seismic applications in multi-parameter full waveform inversion. Numerical results show that our inversion algorithm provides promising results for seismic inversion applicable to realistic model sizes. The 3D numerical test shows that the proposed algorithm can deal with the 3D model (more than 0.6 million unknowns) in an ordinary computer (with an Intel i7-7700 CPU and 64 GB of RAM). This is impossible for the conventional distorted Born iterative method.  
				
				Although we have some favorable results, certain aspects still require attention in future work. Firstly, it is essential to incorporate an absorbing boundary in the numerical test to reduce boundary reflections. However, implementing an absorbing boundary within the integral equation method remains challenging, so it has yet to be included in this study. Secondly, it would be interesting to explore various parametrizations (for example, Thomsen parameters) to assess whether they can help further mitigate the crosstalk issue.

				\section{ACKNOWLEDGEMENTS}
				
							
				\newpage
				\append[op_Green]{2D and 3D elastic Green's function}
				The analytical Green's tensor in a 3D homogeneous isotropic medium is given by \citet{aki1980quantitative} as
				\begin{equation}\label{iner5}
				\begin{aligned}
				G_{i j}(r, \omega)= & \frac{e^{i \omega r / \alpha}}{4 \pi \rho \alpha^2 r}\left[\gamma_i \gamma_j+\left(3 \gamma_i \gamma_j-\delta_{i j}\right)\left(\frac{-\alpha}{i \omega r}\right)\right. 
				\left.+\left(3 \gamma_i \gamma_j-\delta_{i j}\right)\left(\frac{-\alpha}{i \omega r}\right)^2\right]\\
				& -\frac{e^{i \omega r / \beta}}{4 \pi \rho \beta^2 r} \left[ \left(\gamma_i \gamma_j-\delta_{i j}\right)+\left(3 \gamma_i \gamma_j-\delta_{i j}\right)\left(\frac{-\beta}{i \omega r}\right)+\left(3 \gamma_i \gamma_j-\delta_{i j}\right)\left(\frac{-\beta}{i \omega r}\right)^2\right],
				\end{aligned}
				\end{equation} 
				where
				\begin{equation}\label{}
				r=\left|\mathbf{x}-\mathbf{x}^{\prime}\right|, \qquad \gamma_i= \frac{\mathbf{x}-\mathbf{x}^{\prime}}{r} \cdot \mathbf{e}_i, \qquad \delta_{i j}=\mathbf{e}_i \cdot \mathbf{e}_j .
				\end{equation} 
				In a 2D homogeneous isotropic medium, the analytical formula of Green's tensor is given by \citet{sanchez1991diffraction} as
				\begin{equation}\label{}
				G_{i j}(r, \omega)=\frac{1}{i 8 \rho}\left\{A \delta_{i j} - B\left(2 \gamma_i \gamma_j-\delta_{i j}\right)\right\},
				\end{equation} 
				where
				\begin{equation}\label{}
				A=\frac{H_0^{(2)}(\omega r / \alpha )}{\alpha^2}+\frac{H_0^{(2)}(\omega r / \beta)}{\beta^2} \text { and } B=\frac{H_2^{(2)}(\omega r / \alpha)}{\alpha^2}-\frac{H_2^{(2)}(\omega r / \beta)}{\beta^2},
				\end{equation} 
				and $ H_0^{(2)} $ and $ H_2^{(2)} $ are the second kind Hankel functions of 0 order and 2 order, respectively. In the above equations, $ \alpha $, $ \beta $, and $ \rho $ are the homogeneous isotropic background medium's P wave velocity, S wave velocity, and density, respectively.

				\newpage
				\append[op_ad]{Derivation of the adjoint operators}
				The Fr{\'e}chet operator and its adjoint operator satisfy the inner product rule:
				\begin{equation}\label{iner}
					\left\langle\delta \mathbf{u},\mathcal{F} \delta \mathbf{m}\right\rangle_D =\left\langle \mathcal{F}^{\dagger} \delta \mathbf{u},\delta \mathbf{m}  \right\rangle_{\Omega},
				\end{equation}
				where $ \left\langle \cdot, \cdot \right\rangle$ is the inner product on Hilbert space. According to the definition of inner product, we rewrite equation \eqref{iner} as
				\begin{equation}\label{iner2}
					\begin{aligned} 
						& \int d \mathbf{r} \delta \mathbf{u}^{*}\left(\mathbf{r}\right)\left[\mathcal{F}^{(0)} \delta \mathbf{m}^{(0)}\right] \left(\mathbf{r}\right) \equiv \int d \mathbf{x}
						\left[(\mathcal{F}^{(0)})^{\dagger} \delta \mathbf{u}\right]^{*}\left(\mathbf{x}\right) \delta m^{(0)}\left(\mathbf{x}\right), \\	
						& \int d \mathbf{r} \delta \mathbf{u}^{*}\left(\mathbf{r}\right)\left[\mathcal{F}^{(p)} \delta \mathbf{m}^{(p)}\right] \left(\mathbf{r}\right)
						\equiv \int d \mathbf{x}
						\left[(\mathcal{F}^{(p)})^{\dagger} \delta \mathbf{u}\right]^{*}\left(\mathbf{x}\right) \delta m^{(p)}\left(\mathbf{x}\right), \\
					\end{aligned}
				\end{equation} 
				where $ \left[\cdot\right]^{*} $ denotes complex conjugation. Inserting \eqref{F} into the left hand of \eqref{iner2} yields 
				\begin{equation}
					\begin{aligned} 
						& \int d \mathbf{r} \delta \mathbf{u}^{*}\left(\mathbf{r}\right)\left[\mathcal{F}^{(0)} \delta \mathbf{m}^{(0)}\right] \left(\mathbf{r}\right)
						= \int d \mathbf{r} \delta \mathbf{u}^{*}\left(\mathbf{r}\right) \omega^{2} \int d \mathbf{x} \mathbf{G}^{(b)}\left(\mathbf{r},\mathbf{x}\right)  \mathbf{u}^{(b)}\left(\mathbf{x}\right) \delta m^{(0)} \left(\mathbf{x}\right), \\	
						& \int d \mathbf{r} \delta \mathbf{u}^{*}\left(\mathbf{r}\right)\left[\mathcal{F}^{(p)} \delta \mathbf{m}^{(p)}\right] \left(\mathbf{r}\right)
						= -\int d \mathbf{r} \delta \mathbf{u}^{*}\left(\mathbf{r}\right)\int d \mathbf{x} \nabla_{\mathbf{x}} \mathbf{G}^{(b)}\left(\mathbf{r},\mathbf{x}\right) 
						\cdot \mathbf{B}^{(p)} :\nabla_{\mathbf{x}} \mathbf{u}^{(b)} \left(\mathbf{x}\right) \delta m^{(p)} \left(\mathbf{x}\right). \\
					\end{aligned}
				\end{equation} 
				By reordering the integral, we obtain
				\begin{equation}\label{iner3}
					\begin{aligned} 
						& \int d \mathbf{r} \delta \mathbf{u}^{*}\left(\mathbf{r}\right)\left[\mathcal{F}^{(0)} \delta \mathbf{m}^{(0)}\right] \left(\mathbf{r}\right)
						= \int d \mathbf{x} \omega^{2} \int d \mathbf{r} \delta \mathbf{u}^{*}\left(\mathbf{r}\right) \mathbf{G}^{(b)}\left(\mathbf{r},\mathbf{x}\right) \mathbf{u}^{(b)}\left(\mathbf{x}\right) \delta m^{(0)} \left(\mathbf{x}\right),  \\	
						& \int d \mathbf{r} \delta \mathbf{u}^{*}\left(\mathbf{r}\right)\left[\mathcal{F}^{(p)} \delta \mathbf{m}^{(p)}\right] \left(\mathbf{r}\right)
						= -\int d \mathbf{x}  \int d \mathbf{r} \delta \mathbf{u}^{*}\left(\mathbf{r}\right) \nabla_{\mathbf{x}} \mathbf{G}^{(b)}\left(\mathbf{r},\mathbf{x}\right) 
						\cdot \mathbf{B}^{(p)} :\nabla_{\mathbf{x}} \mathbf{u}^{(b)} \left(\mathbf{x}\right) \delta m^{(p)} \left(\mathbf{x}\right). \\
					\end{aligned}
				\end{equation} 
				Comparing the right hand term of \eqref{iner3} and the right hand term of \eqref{iner2}, we have
				\begin{equation}\label{iner4}
					\begin{aligned} 
						& \left[(\mathcal{F}^{(0)})^{\dagger} \delta \mathbf{u}\right] \left(\mathbf{x}\right)
						= \left[\omega^{2} \int d \mathbf{r} \delta \mathbf{u}^{*}\left(\mathbf{r}\right)   \mathbf{G}^{(b)}\left(\mathbf{r},\mathbf{x}\right) \mathbf{u}^{(b)}\left(\mathbf{x}\right)\right]^{*},  \\	
						& \left[(\mathcal{F}^{(p)})^{\dagger} \delta \mathbf{u}\right] \left(\mathbf{x}\right)
						= \left[-\int d \mathbf{r} \delta \mathbf{u}^{*}\left(\mathbf{r}\right) \nabla_{\mathbf{x}} \mathbf{G}^{(b)}\left(\mathbf{r},\mathbf{x}\right) 
						\cdot \mathbf{B}^{(p)} :\nabla_{\mathbf{x}} \mathbf{u}^{(b)} \left(\mathbf{x}\right)\right]^{*}. \\
					\end{aligned}
				\end{equation} 
				Using the reciprocity of the Green's function and moving the differentiation operator on the Green's function out of the integral yields
				\begin{equation}\label{iner5}
				\begin{aligned} 
					& \left[(\mathcal{F}^{(0)})^{\dagger} \delta \mathbf{u}\right]\left(\mathbf{x}\right)
					= \left[\omega^{2} \int d \mathbf{r} \mathbf{G}^{(b)}\left(\mathbf{x},\mathbf{r}\right) \delta \mathbf{u}^{*}\left(\mathbf{r}\right) \mathbf{u}^{(b)}\left(\mathbf{x}\right) \right]^{*},  \\	
					& \left[(\mathcal{F}^{(p)})^{\dagger} \delta \mathbf{u} \right]\left(\mathbf{x}\right)
					= \left[-\nabla_{\mathbf{x}}\int d \mathbf{r} \mathbf{G}^{(b)}\left(\mathbf{x},\mathbf{r}\right)  \delta \mathbf{u}^{*}\left(\mathbf{r}\right) 
					\cdot \mathbf{B}^{(p)} :\nabla_{\mathbf{x}} \mathbf{u}^{(b)} \left(\mathbf{x}\right)\right]^{*}. \\
				\end{aligned}
			\end{equation} 
			
			\newpage
			\append[B]{$ \mathbf{B} $ matrices for a VTI medium} 
			The VTI medium is characterized by five independent elastic parameters: $ C_{11} $, $ C_{33} $, $ C_{55} $, $ C_{66} $, and $ C_{13} $. Their corresponding $ \mathbf{B} $ matrices are shown as follows:
			\begin{equation}
			\begin{aligned}
			&\mathbf{B}^{11}=\left[\begin{array}{cccccc}
			1 & 1 & 0 & 0 & 0 & 0 \\
			1 & 1 & 0 & 0 & 0 & 0 \\
			0 & 0 & 0 & 0 & 0 & 0 \\
			0 & 0 & 0 & 0 & 0 & 0 \\
			0 & 0 & 0 & 0 & 0 & 0 \\
			0 & 0 & 0 & 0 & 0 & 0
			\end{array}\right], 	\quad
			\mathbf{B}^{33}=\left[\begin{array}{cccccc}
			0 & 0 & 0 & 0 & 0 & 0 \\
			0 & 0 & 0 & 0 & 0 & 0 \\
			0 & 0 & 1 & 0 & 0 & 0 \\
			0 & 0 & 0 & 0 & 0 & 0 \\
			0 & 0 & 0 & 0 & 0 & 0 \\
			0 & 0 & 0 & 0 & 0 & 0
			\end{array}\right], \qquad \\
			&\mathbf{B}^{55}=\left[\begin{array}{cccccc}
			0 & 0 & 0 & 0 & 0 & 0 \\
			0 & 0 & 0 & 0 & 0 & 0 \\
			0 & 0 & 0 & 0 & 0 & 0 \\
			0 & 0 & 0 & 1 & 0 & 0 \\
			0 & 0 & 0 & 0 & 1 & 0 \\
			0 & 0 & 0 & 0 & 0 & 0
			\end{array}\right], \qquad
			\mathbf{B}^{66}=\left[\begin{array}{cccccc}
			0 & -2 & 0 & 0 & 0 & 0 \\
			-2 & 0 & 0 & 0 & 0 & 0 \\
			0 & 0 & 0 & 0 & 0 & 0 \\
			0 & 0 & 0 & 0 & 0 & 0 \\
			0 & 0 & 0 & 0 & 0 & 0 \\
			0 & 0 & 0 & 0 & 0 & 1
			\end{array}\right], \\
			&\mathbf{B}^{13}=\left[\begin{array}{cccccc}
			0 & 0 & 1 & 0 & 0 & 0 \\
			0 & 0 & 1 & 0 & 0 & 0 \\
			1 & 1 & 0 & 0 & 0 & 0 \\
			0 & 0 & 0 & 0 & 0 & 0 \\
			0 & 0 & 0 & 0 & 0 & 0 \\
			0 & 0 & 0 & 0 & 0 & 0
			\end{array}\right].
			\end{aligned}
			\end{equation}
	    
				\newpage
				\bibliographystyle{seg}  
				\bibliography{paper3}
				
			
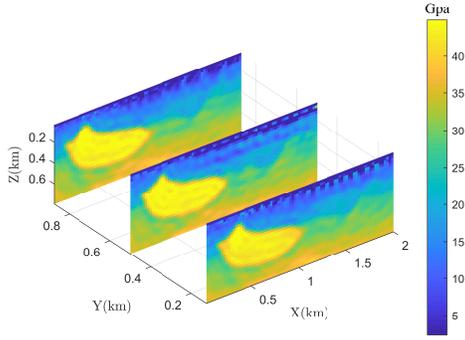
\end{document}